\documentclass[a4paper,fleqn,usenatbib]{mnras}

\usepackage[T1]{fontenc}
\usepackage{ae,aecompl}

\usepackage{amsmath, latexsym, amssymb, graphicx, slashed, color}
\usepackage{hyperref}

\definecolor{nicered}{rgb}{.7,.1,.1}
\definecolor{nicegreen}{rgb}{.1,.5,.1}
\definecolor{darkblue}{rgb}{0,0,.5}
\definecolor{darkblue2}{rgb}{0,0,.7}
\hypersetup{colorlinks, citecolor=nicegreen,linkcolor=nicered, urlcolor=darkblue}

\newcommand{\beq}{\begin{equation}}
\newcommand{\eeq}{\end{equation}}

\newcommand{\gsi}{\,\raisebox{-0.13cm}{$\stackrel{\textstyle>}{\textstyle\sim}$}\,}

\newcommand{\be}{\begin{equation}}
\newcommand{\ee}{\end{equation}}
\newcommand{\bea}{\begin{eqnarray}}
\newcommand{\eea}{\end{eqnarray}}

\newcommand{\bw}{\begin{widetext}}
\newcommand{\ew}{\end{widetext}}

\def\<{\langle}
\def\>{\rangle}

\def\alt{\raise0.3ex\hbox{$\;<$\kern-0.75em\raise-1.1ex\hbox{$\sim\;$}}}
\def\agt{\raise0.3ex\hbox{$\;>$\kern-0.75em\raise-1.1ex\hbox{$\sim\;$}}}

\newcommand\kpc{\text{kpc}}

\newcommand\eV{\text{eV}}

\title[Phase space mass bound from DSph galaxies]{Phase space mass bound for fermionic dark matter from dwarf spheroidal galaxies}

\author[C. Di Paolo et al.]{%
Chiara Di Paolo,$^{1}$\thanks{chiara.dipaolo@sissa.it}
Fabrizio Nesti,$^{2,3,4,5}$\thanks{fabrizio.nesti@irb.hr}
and Francesco L. Villante$^{5,6}$\thanks{francesco.villante@aquila.infn.it}
\vspace*{1.ex}%
\\
$^{1}$SISSA/ISAS, Via Bonomea 265, 34136 Trieste, Italy\\
$^{2}$Dipartimento di Fisica, Theoretical section, Universit\`a di Trieste, Strada Costiera 11, I-34151
Trieste, Italy\\
$^{3}$Ru\dj er Bo\v{s}kovi\'c Institute, Division of Theoretical Physics, Bijeni\v{c}ka cesta 54, 10000, Zagreb, Croatia\\
$^{4}$INFN Sez.\ Trieste, Via A. Valerio 2, 34127 Trieste, Italy\\
$^{5}$Dipartimento di Scienze Fisiche e Chimiche, Universit\`a dell'Aquila, via Vetoio SNC, I-67100, L'Aquila, Italy\\
$^{6}$INFN-LNGS, Via G. Acitelli 22, 67100, Assergi (L'Aquila), Italy%
\vspace*{-3ex}%
}

\begin{document}
\label{firstpage}
\pagerange{\pageref{firstpage}--\pageref{lastpage}}

\maketitle

\begin{abstract}
\noindent
We reconsider the lower bound on the mass of a fermionic dark matter (DM) candidate resulting from
the existence of known small Dwarf Spheroidal galaxies, in the hypothesis that their DM halo is
constituted by degenerate fermions, with phase-space density limited by the Pauli exclusion
principle.  By relaxing the common assumption that the DM halo scale radius is tied to that of the
luminous stellar component and by marginalizing on the unknown stellar velocity dispersion
anisotropy, we prove that observations lead to rather weak constraints on the DM mass, that could be
as low as tens of eV.  In this scenario, however, the DM halos would be quite large and massive, so
that a bound stems from the requirement that the time of orbital decay due to dynamical friction in
the hosting Milky Way DM halo is longer than their lifetime.  The smallest and nearest satellites
Segue I and Willman I lead to a final lower bound of $m\gtrsim100$\,eV, still weaker than previous
estimates but robust and independent on the model of DM formation and decoupling.  We thus show that
phase space constraints do not rule out the possibility of sub-keV fermionic DM.  \vspace*{.2ex}

\hfill{\em to Giulia D.S.}
\end{abstract}


\begin{keywords}
dark matter -- galaxies: dwarf  -- elementary particles -- neutrinos 
\vspace*{0ex}
\end{keywords}

\section{Introduction}\label{introduction}

\noindent
Dark matter (DM) is believed to be a main actor in cosmology, to constitute the great majority of
the mass in the universe and to rule the processes of structure formation.  The so-called
$\Lambda$CDM paradigm, in which one assumes a cold dark matter (CDM) candidate that decouples from
the primordial plasma when non-relativistic, successfully reproduces the structure of the cosmos
down to scales $\sim 50\,$kpc.

A number of serious challenges to the $\Lambda$CDM paradigm have emerged on the scale of individual
galaxies and their central structure, see e.g.~\cite{Weinberg:2013aya} for a recent review.
For instance, collisionless N-body simulations predict that the DM density profile of virialized
objects has a negative logarithmic slope towards the center~\citep{Flores:1994gz, Navarro:1995iw,
  Navarro:1996gj, Moore:1994yx, Moore:1999gc}.  Such a 'cuspy' distribution is not well supported by
observational data of rotation curves of spiral galaxies, which are better described by halos
featuring a constant density core~\citep{Salucci:2000ps}.
Moreover, the number of DM subhaloes expected according $\Lambda$CDM paradigm is much larger than
the observed number of satellite galaxies in the Milky Way \citep{Moore:1999nt, Klypin:1999uc}, even
accounting for the many recently discovered faint systems.
It is still unclear whether the above problems require major changes to the $\Lambda$CDM paradigm.
Models have been presented in which shallow DM cores arise naturally in a $\Lambda$CDM cosmology as a
results of SN feedback or dynamical friction \citep{Navarro:1996bv, Governato:2009bg,
  Governato:2012fa, Pontzen:2011ty}. Alternative DM candidates, however, have to be considered with
utmost attention.

The hypothesis of warm dark matter (WDM) decoupling from the plasma when mildly relativistic,
has been advocated as a solution of the possible CDM issues.  WDM introduces a non vanishing free
streaming length below which structure formation is suppressed, giving rise to the correct abundance
of substructures at small scales \citep{Colin:2000dn, Bode:2000gq, AvilaReese:2000hg}.
Moreover, if we consider a generic fermionic dark matter (FDM) candidate, like the typical massive
$\sim{\rm keV}$ warm sterile neutrino, the limit on the phase space density provided by the Pauli
exclusion principle implies that DM has a minimal velocity dispersion and, thus, resists
compression.
As a consequence, FDM halos naturally produce a cored density profile whose radius $R_{h}$ (for
a fixed halo mass $M_{h}$) is a decreasing function of the mass of $m$ of the DM candidate, see
next section for details.  Being the most DM dominated astrophysical objects, dwarf spheroidal
galaxies are the optimal candidates to test this scenario.

The possibility to constrain the DM particle mass by determining the DM phase space distribution was
first considered in the seminal work by \citet{Tremaine:1979we}.  In the hypothesis of
non-dissipative evolution, i.e.\ conservation of the maximal phase space density, it is possible to
set a strong bounds on the DM mass $m>$300--700\,eV, see e.g.~\cite{Dalcanton:2000hn}.
These bounds stem from the primordial limit of the DM phase space density, and are not necessarily
related to the fermionic nature of dark matter; indeed they apply also if dark matter can be treated
as a collisionless gas collapsed via violent relaxation {\em\`a la} Lynden-Bell.
However, they require the knowledge of the initial DM phase space density (and, thus, of the DM
production mechanism) and the assumption that baryonic feedback cannot alter the maximum of the DM
distribution function.
%
%
A more general situation was studied by~\citet{1992ApJ...389L...9G} that pointed out the importance
of constraints by dynamical friction on dwarf spheroidal galaxies in the Milky Way host halo, that as we
will show still play the most important role.

Subsequent analyses~\citep{Bilic:2001es,Chavanis:2002rj,deVega:2013jfy} modeled truly fermionic DM
cores for instance by using a Thomas-Fermi self-gravitating-gas approach, suggesting that dwarf
spheroidal galaxies host degenerate fermionic halos while larger and less dense galaxies behave as
non-degenerate classical systems.  Observational data on DSph galaxies were used in
\citet{Boyarsky:2008ju} to derive bounds of the order of 400\,eV.  Finally, in a recent
analysis~\citep{Domcke:2014kla} it was claimed that the rotation curves of the eight classical dwarf
spheroidal galaxies of the Milky Way are well fitted by assuming DM cores composed by fully
degnerate fermions with masses $m\simeq 200\,{\rm eV}$ and allowing for a non vanishing anisotropy
of the stellar component.

The observation that kinematic properties of dwarf spheroidal galaxies may be connected, in a
relatively simple model, to the elementary properties of the DM candidate is extremely interesting.
However, there is at the moment no evidence in favour of degenerate galactic cores.
If all the smallest galactic cores were to be degenerate, their masses and radii should follow a
relationship $M_{h} \propto R_{h}^{-3}$, as expected for degenerate fermionic systems and
univocally determined by the mass $m$ of the DM candidate.
This behaviour in the plane $(R_{h},\, M_{h})$ is presently not observed; on the contrary,
the observation that the estimated surface densities of diverse kinds of galaxies is approximately
constant \citep{Donato, Salucci:2011ee}, $\Sigma_0 \sim M_{h}/R_{h}^2 \sim 100\,M_{\odot}
/{\rm pc}^2$, can only be supported by non degenerate cores, because this relation lies almost
orthogonal to the above degeneracy lines.
Still, this argument cannot be used to rule out the existence of degenerate cores, because they
could be present just in the smallest galaxies. These have a larger density, and therefore are
candidate to support or exclude the hypothesis of fermionic DM with low mass.\footnote{It is of
  course also possible that the evolution of structures is such that DM, altough fermionic, never
  forms degenerate cores; in this situation it is still or even more important to assess the values
  of the DM mass that allows the realization of this scenario.}

In this paper we take a conservative attitude and determine a robust lower limit on the mass $m$ of
a fermionic DM particle from the properties of these smallest observed galaxies.
We consider Willman I~\citep{Willman} and Segue I~\citep{Simon:2010ek}, which are among the smallest
structures for which stellar velocity measurements are available, and the ``classical'' dwarf
spheroidal Leo~II from which restrictive bounds on $m$ where obtained by~\citet{Boyarsky:2008ju}
and~\citet{Domcke:2014kla}.
We determine bounds on the core radius $R_{h}$, mass $M_{h}$ or surface density $\Sigma_0$
of the selected galaxies by performing a fit to the stellar line-of-sight velocity dispersion
profile.
The theoretical predictions are obtained through a standard Jeans analysis, including the role of
the unknown velocity dispersion anisotropy of the stellar component.  Moreover, we refrain from
assuming that luminous matter traces the DM distribution, unlike many of the recent works, and thus
leave as a free parameter the DM core radius.

\begin{figure}
\centerline{\includegraphics[width=\columnwidth]{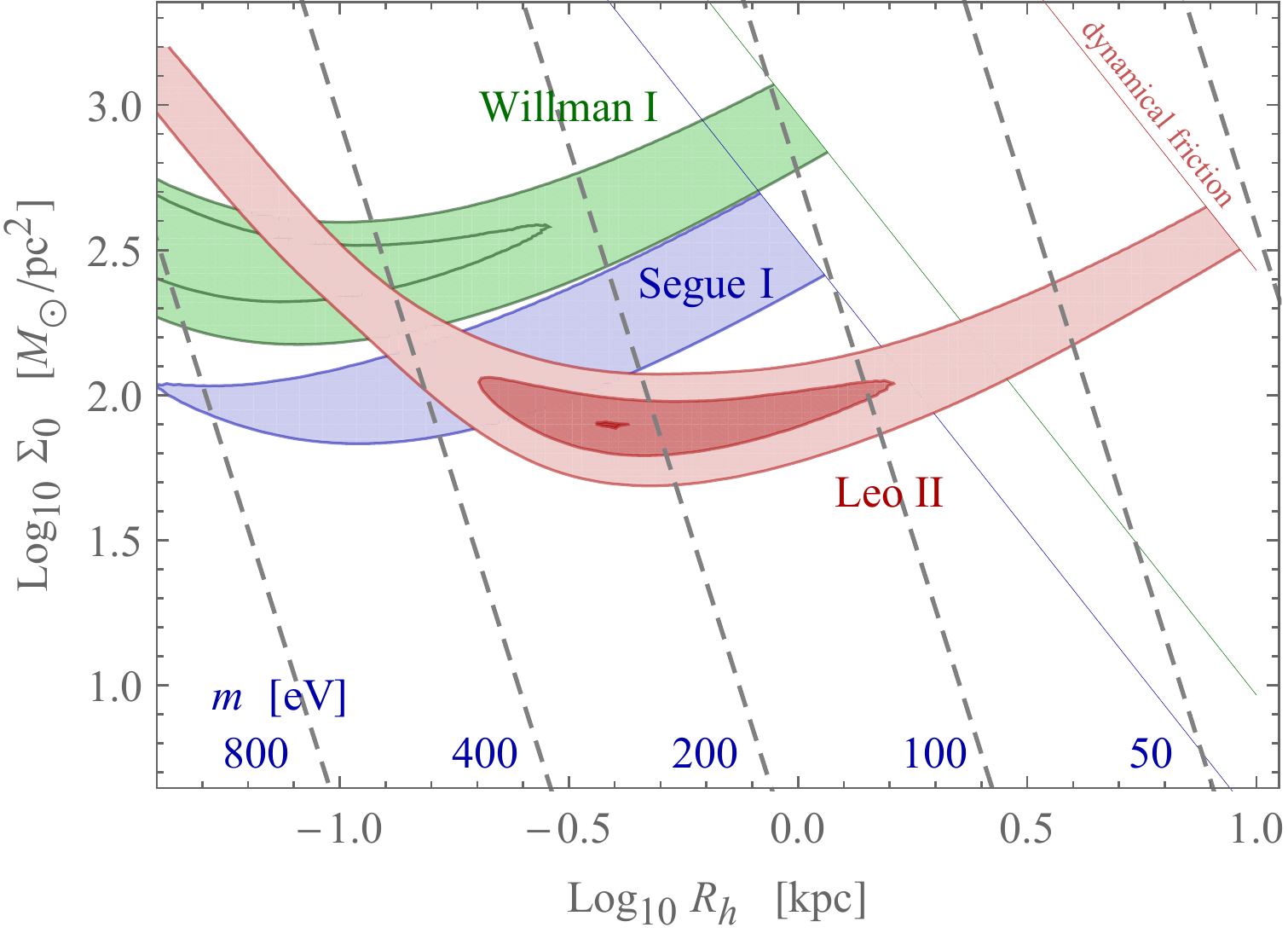}}
\vspace*{-1ex}
\caption{Plane $R_{h}$ -- $\Sigma_0$, describing the DM core. The shaded contours shows the values
     of DM core sizes allowed for Willman I, Segue I and Leo II at 68\% C.L.\ by the LOS velocities,
     and respecting the bound from the dynamical friction time.  \label{fig_bananas}\vspace*{-2ex}}
\end{figure}

We show in detail how, unless the anisotropy of stellar component will be constrained independently,
the observed stellar velocity dispersion profiles lead to very poor constraints on the DM halo, and
the possibility of very large $\sim$kpc halos cannot be ruled out.
However, such large halos are at odds with their lifetime due to the dynamical friction within the
Milky-Way~\citep{BinneyTremaine}.  This provides a further quantitative limit on the DM halo size,
allowing us to finally constrain the DM particle mass $m$.  To facilitate the reader, our results
are anticipated in Figure~\ref{fig_bananas} that contains a synthesis of our work before discussing
the technical details.

The final limit that we obtain, $m\gsi 100\,{\rm eV}$, is less restrictive but more solid than
previously quoted bounds~\citep{Boyarsky:2008ju,Domcke:2014kla} which rely on the assumption that the
DM core radius is equal to the half-light radius, or analogously that the escape velocity from the
DM core is determined by the {\em stellar} velocity dispersion.
Moreover, our limit is fairly model-independent because it is based only on the present phase-space
density of DM particles and does not requires any assumption on their initial distribution or their
evolution (see e.g. \cite{Boyarsky:2008ju} for a discussion of the model-dependent bounds that can
be obtained for a dissipationless DM candidate by considering specific production mechanisms).
Restrictive limits ($m\gsi {\rm few} \,{\rm keV}$) on sterile neutrino mass can be also obtained
from the analysis of the Ly-$\alpha$ forest data, see e.g.~\cite{Irsic:2017ixq} for a recent update.
It should be noted, however, that this analysis is not directly sensitive to DM particle mass, as
Ly-$\alpha$ data essentially probe the power spectrum of density fluctuations at comoving scales
$\sim{\rm Mpc}$, by constraining the DM free streaming length. Since this quantity can be related to
the particle mass only within a specific DM model, the Ly-$\alpha$ bound cannot be applied to a
generic fermionic candidate, unlike from the limit derived in this paper.

The plan of the paper is the following: in section~\ref{sec:fwdm} we describe the physics relevant
for degenerate fermionic dark matter halos, while in section~\ref{sec:strategies} we lay down the
possible strategies to constrain the mass of the dark matter candidate from the observational
data. In section~\ref{sec:results} we describe the results for the Leo II, Willman I and Segue I
dwarf galaxies, by paying also attention to the possibility that some of these galaxies are instead
non-degenerate. We present also a consistent estimate for the ensuing bound on the DM mass in the
case of the other known dsph galaxies. In section~\ref{sec:conclusions} we summarize the conclusions
and possible outlook. Finally, for convenience we briefly review in Appendix~\ref{app:review} the
technicalities relative to degenerate fermionic halos, as well as the Thomas-Fermi analysis for non
exactly degenerate ones.

\section{The FDM hypothesis}
\label{sec:fwdm}

\noindent
We consider the equilibrium configuration for an ensemble of self-gravitating DM fermions of mass
$m$ and $g$ internal (spin) degrees of freedom. The assumption of fermionic particles implies the
upper limit for the DM phase space distribution function that, as reviewed in
appendix~\ref{app:review}, translates into a lower limit for the DM velocity dispersion
%
\begin{eqnarray}
\sigma^2_{\rm DM}  & \ge &    \sigma^2_{\rm DM, min}(\rho)  
 = \frac{1}{5} \left(\frac{ 6\, \pi^2 \hbar ^3 \rho}{g\, m^4}\right)^{2/3}  \\
\nonumber
& = & 7.56 \, \left(\frac{\rm km}{\rm s}\right)^2 \left(\frac{g}{2}\right)^{-2/3} \left(\frac{m}{\rm 1\, keV}\right)^{-8/3}
\left(\frac{\rho}{\rm M_\odot/ pc^{3}}\right)^{2/3}
\label{eq:sigmaMin}
\end{eqnarray}
as a function of the mass density $\rho$ of the system. This bound becomes effective and very
stringent in the regions of high density.  As a consequence, fermionic DM halos resist compression
and cannot have arbitrary size.

The strong degeneracy limit, in which the velocity dispersion is assumed to have the minimal value
$\sigma^2_{\rm DM, min}(\rho)$, represents the most compact configuration for a self-gravitating
fermionic halo.  The density profiles of such fully degenerate halos are universal.  They depend only
on one free parameter and can be expressed (apart from a normalization factor and a scale radius
which are related, see the following) in terms of the solution of the well-known Lane-Emden
equation, see equation~(\ref{FDM-LaneEmden}).
As shown in the appendix, for our purposes the degenerate profiles are very well approximated by the
function:
\begin{equation}
\rho(r)=\rho_0 \cos^3 \left[\frac{25}{88} \,\pi \,  x\right]
\,,\qquad x= r/R_{h}\,,
\label{DMdensityApp}
\end{equation}
where $\rho_0$ is the central DM halo density. 
The halo radius $R_{h}$ is defined by the condition
\begin{equation}
\rho\left(R_{h}\right) = \rho_0 / 4
\end{equation}
and is related to the central density $\rho_0$ and to the properties of the DM particle by the
numerical relation
\begin{equation}
\label{RRh}
R_{h} =  
42.4 \, {\rm pc}\, \left(\frac{g}{2}\right)^{-1/3}
\left(\frac{m}{\rm 1\, keV}\right)^{-4/3}
\left(\frac{\rho_0}{\rm M_\odot/pc^{3}}\right)^{-1/6}\,.
\end{equation}
This value represents also the minimal admissible radius for a fermionic structure since for smaller
radii the gravitational potential $\phi \sim - G\, \rho_0 \, R_{h}^2$ is lower (in modulus) than
$\sigma^2_{\rm DM, min} \sim (\rho_0/g)^{2/3} \, m^{-8/3}$ and the system is not stable.

Larger non degenerate structures are admissible because they can have $\sigma_{\rm DM}^2\ge
\sigma^2_{\rm DM, min}$ that prevents gravitational collapse.  Unlike in the completely degenerate
case, their properties cannot be univocally predicted because the velocity dispersion is not
determined by the mass density and not directly linked to the DM particle properties.
Isothermal halos with arbitrary level of degeneration can be studied by using the Thomas-Fermi
approach as reviewed in Appendix~\ref{app:review}.  Interestingly, it is found that when $R_{h}$
is just 2--3 times larger than the minimal value~(\ref{RRh}), the fermionic nature of DM particles
can be neglected, i.e.\ the resulting structures are essentially indistinguishable from isothermal
halos obtained by assuming classical Maxwell-Boltzmann statistics and arbitrary large values of the
particle mass $m$.

For fully degenerate fermionic structures, by using equation~(\ref{RRh}) one can predict their surface
density $\Sigma_0\equiv \rho_0\, R_{h}$
\begin{equation}
\label{SSigma0}
\frac{\Sigma_0}{\rm M_\odot/pc^{2}} = 0.584 \,
\left(\frac{g}{2}\right)^{-2}
\left(\frac{m}{\rm 1\, keV}\right)^{-8}
\left(\frac{R_{h}}{\rm 100\, pc}\right)^{-5}
\end{equation}%
as well as their mass $M_{h}$, defined as the mass enclosed within the radius $R_{h}$:
\begin{equation}
\label{MMh}
\frac{M_{h}}{\rm 10^{7} M_\odot} = 1.18\,  
\left(\frac{g}{2}\right)^{-1/3}
\left(\frac{m}{\rm 1\, keV}\right)^{-8}
\left(\frac{R_{h}}{\rm 10\, pc}\right)^{-3}\,.
\end{equation}
The radius, surface density and mass of degenerate halos are not independent quantities, being
$M_{h} \simeq 2.02 \; \rho_0 \, R_{h}^3 = 2.02 \; \Sigma_0 \, R_{h}^2$ for the
density profile~(\ref{DMdensityApp}).  For definiteness, we perform our analysis in the plane
$(R_{h} ,\, \Sigma_{0})$ but equivalent bounds are clearly obtained by using any couples of the
three quantities $R_{h}$, $\Sigma_0$ and $M_{h}$.

In Figure~\ref{fig_bananas} we have reported as gray dashed lines in the plane $(R_{h} ,\,
\Sigma_{0})$ the points relative to fully degenerate systems for selected values of the DM particle
mass $m$ and assuming $g=2$.  Eqs.~(\ref{SSigma0}) and (\ref{MMh}) define the lower limits for
surface densities and masses of fermionic DM halos.
The regions to the left of the gray lines in the plane $(R_{h} ,\, \Sigma_{0})$, are not
compatible with the assumption that the halo is composed by fermionic particles.
Note that for fixed surface density, the smaller is the particle mass the larger has to be the core
radius.
As a consequence, the observational determinations of halo radii $R_{h}$ and surface densities
$\Sigma_0$ can be translated into lower limits for the mass $m$ of fermionic dark matter candidates:
\begin{equation}
\frac{m}{\rm keV} \ge 0.53\,
\left(\frac{g}{2}\right)^{-1/4}
\left(\frac{\Sigma_0}{100 \, \rm M_\odot/pc^{2}} \right)^{-1/8}
\left(\frac{R_{h}}{\rm 100\,pc}\right)^{-5/8}
\label{eq:mbound}
\end{equation}
and we note the reduced dependence on $\Sigma_0$.

If galactic cores were to be commonly degenerate, one should arguably observe a clustering along the
lines defined by equation~(\ref{SSigma0}) in the plane $(R_{h} ,\, \Sigma_{0})$, at least for the
smallest structures.
This clustering is presently not observed.
Instead, the estimated surface densities of diverse kinds of observed galaxies appear to be
approximately constant \citep{Donato, Salucci:2011ee}, $\Sigma_0\sim 100 M_\odot / {\rm pc}^2$, a
fact that can only be supported by non degenerate cores, because this relation lies almost
orthogonal to the degeneracy lines.

This argument cannot be used to rule out the fermionic nature of DM, or the occurrence of degenerate
halos, because even if all large galaxies host a non-degenerate DM halo, degenerate cores could be
present in the smallest objects, of limited number and maybe even too small to be observed.
Therefore, what one can do at present is to obtain a lower limit on the mass of a fermionic DM
candidate from the existence of the smallest galaxies, once their properties (radius, mass and/or
surface density) are determined. It is our aim to reassess in this way the present bound on $m$.

\section{Strategies}
\label{sec:strategies}

\noindent
The standard mass estimation methods applied to dwarf spheroidal galaxies, like those described
in~\cite{Wolf}, are not sufficient for our purposes.  In fact, they provide an estimator of the mass
$M_{1/2}$ enclosed inside the half-light radius $R_{1/2}$, but this radius is not necessarily
representative of the DM distribution and might be in principle (much) smaller than the halo
size $R_{h}$.  In other words, the quantity $M_{1/2}$ represents only a lower limit on the core
mass $M_{h}$ but does not provide an upper constraint.

In e.g.~\cite{Boyarsky:2008ju} and \cite{Domcke:2014kla}, lower bounds on the mass of FDM
candidates were obtained by assuming that $R_{h}\simeq R_{1/2}$, i.e.\ $M_{h}\simeq M_{1/2}$
and/or by assuming that the stellar velocity dispersion can be used to estimate the escape velocity
from the DM core.
This is only allowed if we assume that luminous matter traces DM distribution.
However, this assumption may well be violated, especially in the considered scenario in which the
properties of the DM distribution are not only determined by gravitational interactions but also by
the fermionic nature of DM.
One may also recall that for larger galaxies, like the Milky Way or elliptical galaxies, the scale
lengths of stellar and dark components can differ greatly, with the dark component extending
typically some factors more than the stellar one.

Along these arguments, in this work we proceed in more generality, treating $R_{h}$ and $M_{\rm
  h}$ as independent properties of the DM halo.  We only use $R_{1/2}$ and $M_{1/2}$ in a
preliminary stage to select, by using the values tabulated in \cite{Wolf} and equation~(\ref{MMh}), the
dwarf spheroidal galaxies Willman~I \citep{Willman} and Segue~I~\citep{Simon:2010ek} as the most
promising candidates for constraining $m$.
In addition to these two galaxies, which are among the smallest structures for which stellar
velocity measurements are available, we also consider the ``classical'' dwarf spheroidal Leo II
from which restrictive bounds on $m$ where obtained by~\citet{Boyarsky:2008ju} and \citet{Domcke:2014kla}.

For each galaxy, we determine the DM halo properties, core radius and mass (or surface density) by
performing a fit of the stellar line-of-sight velocity dispersion profile as predicted by the model
through the Jeans analysis, to the observed data. We also consider the role of the possible stellar
velocity dispersion anisotropy. As was already suggested in the past (see
e.g.~\cite{2013pss5.book.1039W}) in most cases the poor data and the unknown anisotropy lead to very
poor constraints on the DM halo, and the possibility of very large $\sim $kpc halo can not be ruled
out.  We then consider that such large haloes would be associated with unphysically short orbital
lifetimes due to the dynamical friction within the Milky-Way. This fact provides a further
quantitative constraint on the DM halo size, and thus allows us to constrain the DM particle mass
$m$.
\vspace*{-1ex}

\subsection{Spherical Jeans analysis}

\noindent 
Assuming that the stellar component is virialized within the gravitational potential
dominated by the DM component, the standard spherical Jeans equation
\be
\left(\frac{\partial }{\partial r}+ \frac{2\beta}{r}\right) (n_*\sigma_r^2)=-n_*\frac{GM(r)}{r^2}\,
\label{eq:jeans}
\ee
allows one to relate the velocity dispersion profile of stars to the DM mass distribution $M(r)$.
In the above, $G$ is the Newton constant, $n_*(r)$ is the stellar number density, $\sigma^2_r$ is
the radial velocity dispersion of stars, and $\beta\equiv 1-\sigma^2_\perp/\sigma^2_r$ is its
anisotropy, which in principle can depend on radius. We first discuss the case of zero anisotropy,
and later comment on its role.
Our final results are obtained by treating $\beta$ as a nuisance parameter.
A number of other aspects, like the possible coexistence of more than one stellar component, or
non-complete virialization, are further factors of uncertainty that may not be easily removed.

We model the density profile of the stellar component for each dwarf spheroidal galaxy by means of a
Plummer density profile with specific scale radius $R_*$:
\be
n_*(r)=n_0\,\left(1+x^2\right)^{-5/2}\,,\qquad x= r/R_*\,,
\label{eq:plummer}
\ee
and the central density $n_0$ plays no role in the following. Clearly, the applicability of this
density profile to real and poorly known galaxies is an other element of uncertainty.

\begin{figure}
\centerline{\includegraphics[width=\columnwidth]{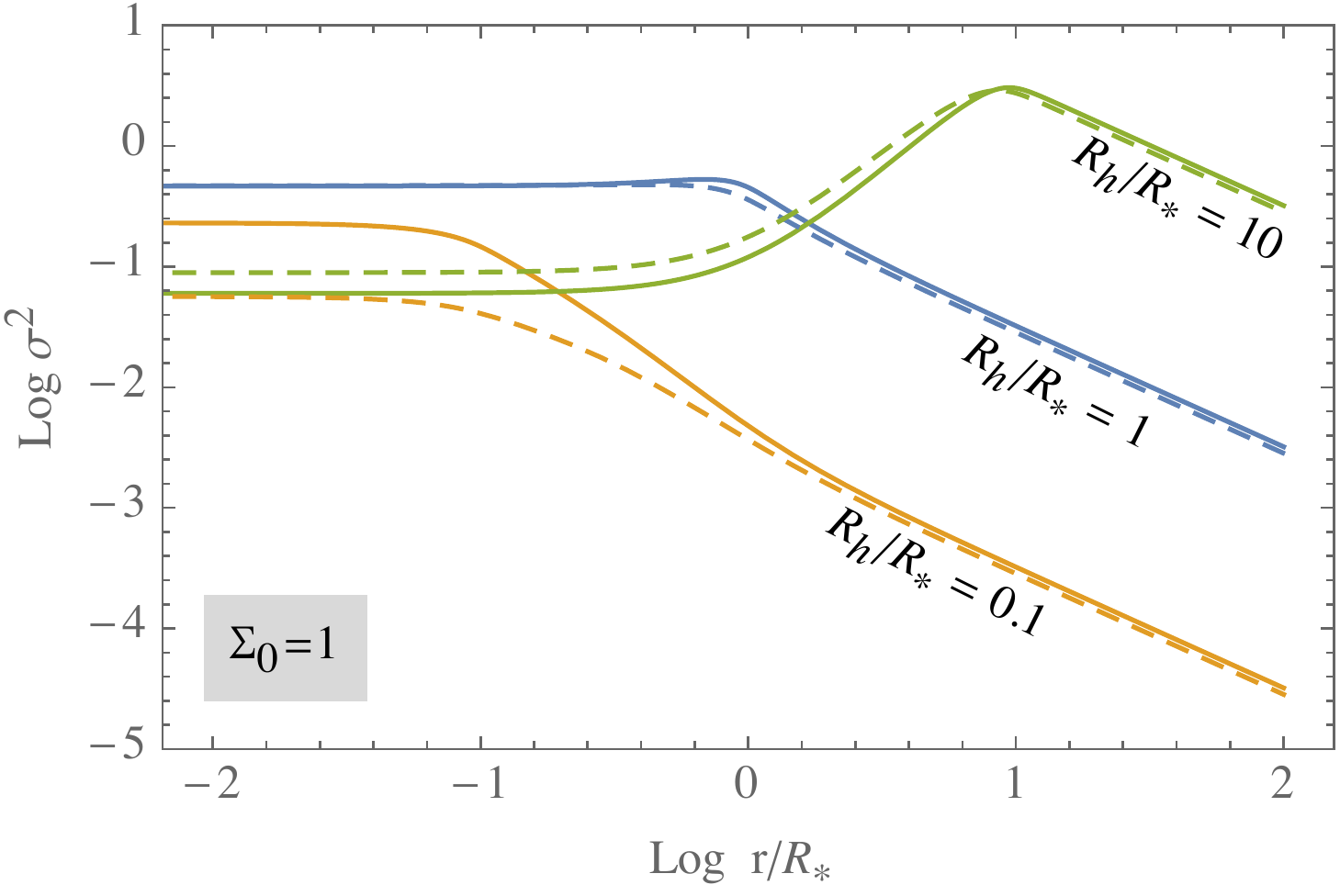}}%
\vspace*{-1ex}
\caption{Stellar velocity dispersion profiles (solid) for representative DM core radii and
  $\beta=0$.  The dashed curves show the line-of-sight projected dispersion velocity
  profiles.\label{fig_stellar_dispersions}}%
\vspace*{-1ex}
\end{figure}

Equation~(\ref{eq:jeans}) can be integrated in favor of $\sigma_r^2$, once the dark matter mass
distribution $M(r)$ is determined by the DM density equation~(\ref{DMdensityApp}).
The resulting stellar velocity dispersion is shown in Figure~\ref{fig_stellar_dispersions}, for 
three representative cases of $R_{h}$ smaller, equal or larger than the stellar scale radius
$R_*$.
The profiles shown are illustrative and are obtained by normalizing to a fixed surface density $\Sigma_0 =
\rho_0 R_{h}=1$. In fact, once the radius $R_{h}$ is fixed, the DM central density $\rho_0$ or the
surface density $\Sigma_0$ represent just a multiplicative constant factor for the mass function
$M(r)$ and do not affect the radial dependence of $\sigma_r^2$. 

We see from Figure~\ref{fig_stellar_dispersions} that if the DM halo is smaller than the Plummer
radius, $R_{h}\le R_{*}$, the stellar velocity dispersion starts to fall as soon as the DM
density vanishes, reflecting the decrease of the gravitational potential.
On the other hand if the DM distribution is more extended than the stellar one, $R_{h}\ge
R_{*}$, the stellar velocity dispersion has to increase in the regions where the Plummer density
drops.
In few words, the slope of the stellar velocity dispersion $\partial\ln \sigma^2_r /\partial\ln r $
is related to the characteristic sizes $R_{*}$ and $R_{h}$ of the galactic components and could, thus,
be used to constrain the DM distribution.

In order to compare with observational data, one has also to consider that only the velocity dispersion
\emph{along the line of sight} (LOS) is measurable:
\be
\sigma_{\rm los}^2(R)=\frac{1}{\Sigma_*}\int_{R^2}^\infty \!{\rm    d}r^2\,
\frac{n_* }{\sqrt{r^2-R^2}}\, \sigma_r^2
\left[1-\beta\frac{R^2}{r^2}\right],
\label{eq:sigmaobs}
\ee 
where $\Sigma_*(R)=\int_{R^2}^\infty {\rm d}r^2 \; n_*(r) /\sqrt{r^2-R^2}$ is the projected stellar
(surface) density. In Figure~\ref{fig_stellar_dispersions}, we show with dashed lines the LOS velocity
dispersion for the three cases described previously. They retain the same behaviour of
$\sigma^2_{r}$, showing that the observed LOS velocity dispersion profile $\sigma_{\rm los}^2(R)$
can in principle be used to constrain the size of the DM core.

\subsection{Analysis for the smallest objects}
\label{sec:aver}

\noindent 
In order to obtain the most restrictive bounds on the mass of the FDM particle, below we
will consider Willman I~\citep{Willman} and Segue I~\citep{Simon:2010ek} which are among the smallest
observed galactic structures.
The problem with these objects is that the number of stars for which a measure of velocity is
available and which pass quality cuts is quite small (i.e. less than 50). As a consequence, it is
not possible to determine the velocity dispersion in radial bins sufficiently localized to be
compared directly with the profile in equation~(\ref{eq:sigmaobs}). One can still appreciate the
characteristic signature of a large DM core, i.e.\ an increasing slope of the
stellar dispersion velocity between $r\sim R_{*}$ and $r\sim R_{h}$, by determining the velocity
dispersion in few relatively large bins with dimension $\Delta r \sim R_*$.

In order to perform such analysis, we define the \emph{average} LOS velocity 
dispersion in a bin $r\in\left[R_1,R_2\right]$,
\be
\overline{\sigma^2_{\rm los}}(R_1,R_2)=\frac{1}{N_{*}(R_1,R_2)}\int_{R_1}^{R_2}{\rm d}r \, 2\pi r \, \Sigma_*(r)\, \sigma^2_{\rm los}(r)\,,
\label{eq:sigmaav0}
\ee
where $N_*(R_1,R_2)=\int_{R_1}^{R_2}{\rm d}r \; 2\pi r \, \Sigma_*(r)$ is the cumulative number of
stars between two radii. 
 Using (\ref{eq:sigmaobs}), and assuming constant $\beta$, we find
\be
\overline{\sigma^2_{\rm los}}=\frac{1}{N_{*}(R_1,R_2)}
\int_{R_1}^{\infty} {\rm d}r \; 4\pi r^2 \, n_* \sigma^2_r \;
\mathcal{F}\left(r,\beta ;R_1, R_2 \right),
\label{eq:sigmaav1}
\ee
where the adimensional function $\mathcal{F}\left(r,\beta;R_1,R_2\right)$ is
\begin{eqnarray}
\nonumber
\mathcal{F}\left(r,\beta;R_1, R_2 \right) &=&
\left\{
\left[\sqrt{r^2-R_1^2}-\sqrt{r^2-B^2}\right]
\frac{\left(3-2\beta\right)}{3 r} + \right.\\
& &
\hspace{-1.3cm}+\left.
\frac{\beta}{3 r^3}
\left[
B^2 \sqrt{r^2-B^2} -
R_1^2 \sqrt{r^2-R_1^2}
\right] \right\}
\end{eqnarray}
with $B=\min\left\{r,R_2\right\}$.\footnote{The number of stars in a bin $N_*(R_1,R_2)$ can be also
  calculated as $N_{*}(R_1,R_2) = \int_{R_1}^{\infty} {\rm d}r \; 4\pi r^2 \, n_*(r) \,
  \mathcal{F}\left(r,\beta=0; R_1, R_2 \right)$.}

\begin{figure}\includegraphics[width=\columnwidth]{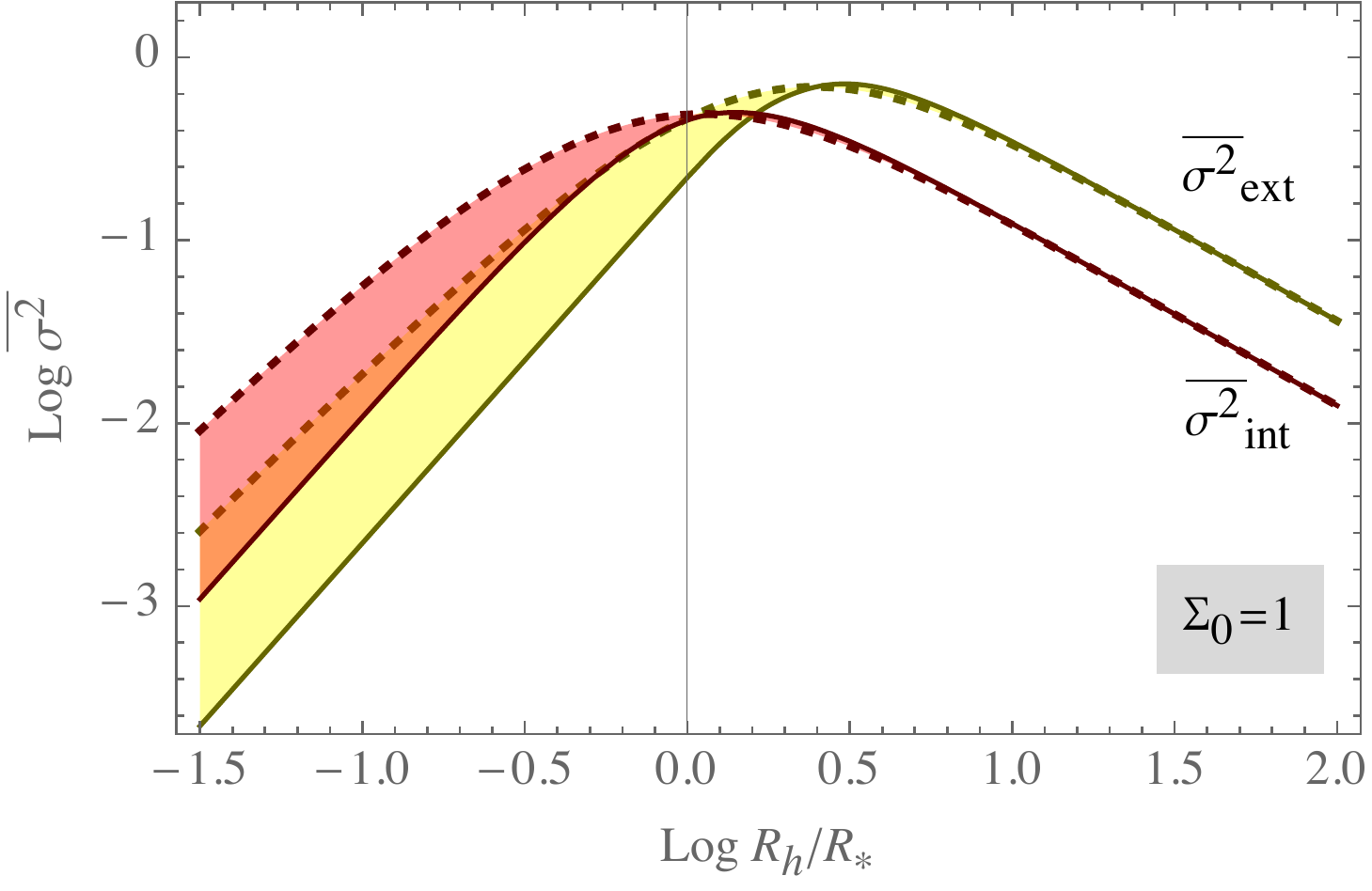}%
\vspace*{-2ex}
\caption{Averaged stellar velocity dispersion in two bins, taken here as $int=[0,r_*]$ (red) and
  $ext=[r_*,3r_*]$ (yellow), for $\beta=0$.  The dashed curves show the same for a Burkert profile
  (non-degenerate fermions).\label{fig_sigma_av}}
\end{figure}

For illustrative purposes, we show in Figure~\ref{fig_sigma_av} as solid lines the behaviour of the
average LOS velocity dispersion calculated in two bins $\left[0,R_*\right]$ and $\left[R_*,3
  R_*\right]$ as a function of the DM core radius $R_{h}$, for $\beta=0$.  One can observe that
the average LOS dispersion velocity in the external bin overshoots the internal one as soon as
$R_{h}/R_*\gtrsim 2$.  This demonstrates that even with two single bins, and provided the
uncertainties on the observed dispersion velocity are not too large, one could constrain the DM core
size. For instance, if the observed dispersion velocities in two or more bins in the vicinity of
$r\simeq R_{*}$ are approximately the same, one could rule out the possibility that DM extends much
beyond the stellar component. 

From the plot one can make also other remarks. Clearly, the more the DM core extends beyond the
stellar component ($R_{h}/R_*>1$) the less its actual density profile beyond $R_{h}$ is relevant for the
stellar physics, because the DM density is anyway constant in the region where the stars trace the
gravitational potential. On the other hand, one can expect that if the DM core is smaller than the
stellar scale ($R_{h}/R_*<1$) the actual shape of the DM profile out of its core will influence the
resulting stellar velocity dispersion. To show this effect, we have repeated the analysis for
Burkert DM profiles, also reported in Figure~\ref{fig_sigma_av} as dashed lines, which confirms the
dependence on the profile shape for $R_{h}<R_*$.  On the other hand, for $R_{h}>R_*$, the solid and
dashed curves are overlapping, i.e.\ the predicted dispersion does not depend on the shape of the DM
profile, making the analysis more robust.

Unfortunately, as we shall discuss for the specific cases of Segue I and Wilman I, the analysis
outlined in this section is considerably hampered by the large uncertainties in each bin of the
observed velocity dispersion. In addition, it is also severely limited by the unknown velocity
anisotropy.

\subsection{The role of anisotropy}
\label{sec:anisotropy}

\noindent 
As is well known ad as we will see in detail, the unknown stellar velocity dispersion anisotropy
$\beta$ limits severely the possibility to extract the DM core radius $R_{h}$ from observational
data.
Indeed, in the absence of direct information, the quantity $\beta$ has to be treated as a nuisance
parameter that has to be removed in order to compare a mass model with observations (see
e.g.~\cite{Walker} and \cite{Ullio:2016kvy} for a recent critical discussion).
The role of $\beta$ can be understood by rewriting the Jeans equation~(\ref{eq:jeans}) in
the form
\be
\frac{\partial\ln \sigma^2_r}{\partial\ln r} =
-\frac {1 }{\sigma^2_r}
\frac {GM }{r} - \gamma_* -2\beta
\label{eq:jeans2}
\ee
where $\gamma_*= d\ln n_*/d\ln r$ is the slope of the stellar number density, that runs from $\sim
0$ near the center to negative values out of $R_*$.
Note that the first two terms in the r.h.s.\ which are related to DM and stellar distributions have
opposite signs, being negative the first and positive the second.
In the case of zero anisotropy, the slope of the velocity dispersion vanishes at the galactic center,
i.e. $\partial\ln \sigma^2_r /\partial\ln r = 0$ for $r=0$, since these two terms are both equal to
zero in the origin.
If the DM halo extends outside the stellar scale radius, the term $\gamma_*$ starts to grow (in
modulus) at $r\simeq R_*$ while $G M(r)/ r \sigma^2_r$ is still negligible, and determines the
positive slope of $\sigma^2_r$ observed in Figure~\ref{fig_stellar_dispersions}. Whereas the observed
stellar velocity dispersion does not feature such a growth at large distances, one can set an upper
bound on the DM core radius, that can not be much larger than the stellar scale radius $R_*$. 

The presence of a non vanishing anisotropy can clearly alter this scenario. In particular, a
positive $\beta$ can compensate the effect of $\gamma_*$ reducing the outer slope of $\sigma^2_r$,
even in presence of a DM distribution extending well beyond the stellar radius. However, this effect
is limited by the fact that the anisotropy parameter is limited to be less than 1. As a result, the
case $\beta=1$ leads to weaker upper bound on the core radius $R_{h}$, and basically the most
conservative.
The opposite holds for negative $\beta$ values that can give quite small slope $\partial\ln
\sigma^2_r /\partial\ln r \ge 0$ even for mass models with $R_{h}\le R_{*}$. Because a negative
$\beta$ is unconstrained, this limits the possibility of setting a lower bound on $R_{h}$ in
dwarf spheroidals and assessing the cusp-core problem at such short scales.

Note that, for the purpose of determining an upper limit for the core radius $R_{h}$, it is not
even necessary to consider a generic $\beta(r)$ since the most relevant fact is the presence of the
upper bound $\beta<1$.
In our analysis, we assume a \emph{constant} anisotropy parameter $\beta$ and we treat it as a
nuisance parameter. 

In order to do this, we adopt the following philosophy: 
for each set of DM core radius $R_{h}$ and surface density $\Sigma_0$ we scan over the complete
physically allowed range $-\infty \le \beta \le 1$, and if there exists a value of $\beta$ that
provides a good fit to the observational data, then the hyphotesis of a degenerate core with those
parameters cannot be rejected. 
Following a frequentist approach, the level of compatibility with data is assessed by defining a
standard $\chi^2$ function, see section~\ref{sec:leoII}. This is eventually minimized with respect
to $\beta$. This procedure allows us to obtain the most conservative bounds.

The comprehensive range in $\beta$ that we adopt, reflects our present ignorance of the dispersion
anisotropy. Clearly, if in the future independent constraints on the anisotropy parameter will
become available, this procedure might be improved.

It should be noted that an additional source of uncertainty is due to the slope of the density
profile $\gamma^*$. As a general rule (see equation~(\ref{eq:jeans2})) the role of $\gamma^*$ in the
outer regions is similar to that of $\beta$: the more negative $\gamma^*$ is, the stronger is the
rise in the external $\sigma_r$ and thus the more restrictive the bound would be.
The Plummer profile that we adopt reaches quite steep values ($-5$) of the slope in the outer region
and thus can be taken as a reasonable benchmark.
Because in practice the precise slope of the density profile is extremely hard to constrain from
observations, especially for the smallest objects, this additional uncertainty should be kept in
mind.

\subsection{Dynamical friction}

\noindent 
The mass of dwarf spheroidals can be limited from above because they are subject to dynamical
friction in the Milky-Way DM halo. Their orbit decay with a characteristic time scale that can be
estimated from the Chandrasekhar's formula~\citep{BinneyTremaine}
\be
\label{tfric}
t_{\rm fric}=\frac{10^{10}\, {\rm y}}{\ln \Lambda} 
  \left(\frac{D}{60\,\kpc}\right)^{\!2}  
  \left(\frac{v}{220\,{\rm km}/{\rm s}}\right)  
  \left(\frac{2\cdot 10^{10}\, M_\odot}{M_{h}}\right)
\ee
where $v$ is the velocity of the dwarf galaxy and $D$ is its distance from the Milky-Way center.
The Coulomb logarithm in the above equation is given by
\be
\ln \Lambda = \ln\left(\frac{b_{\rm max}}{b_{\rm min}}\right),
\ee
where $b_{\rm max}$ and $b_{\rm min}$ are the maximum and minimum impact parameters. These can be
estimated as \citep{BinneyTremaine, 2011MNRAS.411..653J}:
\be
  b_{\rm max} = - \left(\frac{d\ln \rho_{MW}}{dr}\right)^{-1}\!\! \simeq
                \frac{D}{\gamma}\,,\quad 
b_{\rm min} = \max\left\{R_{h}\,, \frac{G M_{h}}{v_{\rm typ}^2}\right\},
\ee
%
where $v_{\rm typ}$ is the virial velocity and we assumed that the Milky Way DM density scales
approximately as $\rho_{\rm MW} \propto D^{-\gamma}$ with $\gamma \simeq 2$ in the vicinity of the
objects considered.\footnote{For degenerate cores, the halo radius $R_{h}$ defined in
  equation~(\ref{eq:rhoDM}) is sufficiently close to the half-mass radius of the DM distribution.}

Chandrasekhar's formula (\ref{tfric}) is known to fail when the mass of the mass $M_{h}$ of the
satellite becomes comparable to the mass of the host system that lies interior to the satellite's
orbit and/or the density of host system is constant, see e.g.~\cite{Read:2006fq}. In the cases of
our interest, however, none of these conditions apply and equation~(\ref{tfric}) provides a remarkably
accurate description.
%
By requiring $t_{\rm fric}\gtrsim 10^{10}y$, and by considering that the typical velocity of
satellites should be of the order of the Galactic virial velocities at those distances $\sim
220\,$km/s~\citep{Nesti:2013uwa}, one finds a bound on the mass $M_{h}$ that depends on the
distance of the dwarf galaxy from the Galactic center.

Note that the existence of an upper limit for $M_{h}$ does not imply by itself the possibility
to constraints the FDM scenario. It was noted, however, in \citet{1992ApJ...389L...9G} 
that if the DM density of dwarf spheroidal galaxies can be determined from velocity dispersion data,
the upper bound on $M_{h}$ can used to obtain an upper limit on $R_{h}$, thus constraining
the mass $m$ of hypothetical FDM particles.

\section{Results}
\label{sec:results}

\subsection{A small classical Dwarf - Leo II}
\label{sec:leoII}

\begin{figure}
  \includegraphics[width=\columnwidth]{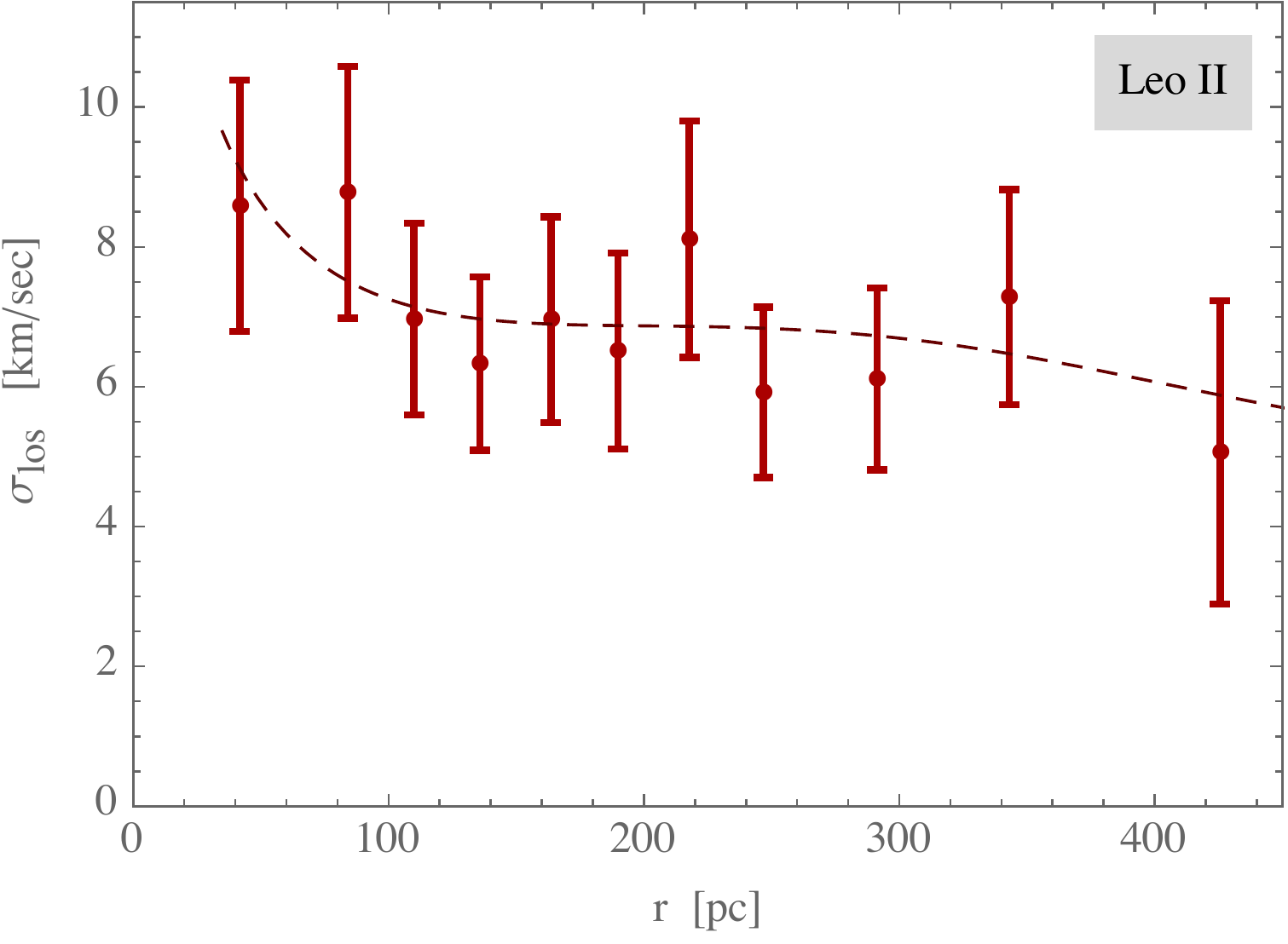}%
  \caption{Stellar line-of-sight velocity dispersions for Leo II. The dashed line represents the
    best fit, achieved for $\beta=0.6$.\label{fig_leo_velocities}}
\end{figure}

\noindent
As a paradigmatic case, we first analyze the case of Leo~II, the smallest among the so-called 'classical' dwarf spheroidal satellite galaxies of the MW.
The stellar number density of Leo~II is well modeled by a Plummer profile with scale length $R_*=177 \,{\rm pc}$. In Figure~\ref{fig_leo_velocities} we report the observed stellar LOS velocity dispersion, $\sigma_i^2\pm\delta\sigma_i^2$, measured in 11 bins centered at the radii $r_{i}$.
We compare these data with the LOS velocity dispersion predicted in
(\ref{eq:sigmaobs}) for the fully degenerate fermionic DM halo, $\overline{\sigma_i^2}\equiv\sigma_{\rm los}^2(r_i)$, by defining a standard $\chi^2$ test:
\be
\chi^2(R_{h},\Sigma_0,\beta) = \sum_{i\in {\rm bins}}\left(\frac{\sigma_{i}^2-\overline{\sigma^2}_i}{\delta \sigma^2_i}\right)^{\!2}.
\label{eq:chi2}
\ee
The model parameters are the DM core radius $R_{h}$ and surface density $\Sigma_0\equiv\rho_0\,R_{h}$,
plus the anisotropy $\beta$.

\begin{figure*}
\includegraphics[width=.95\columnwidth]{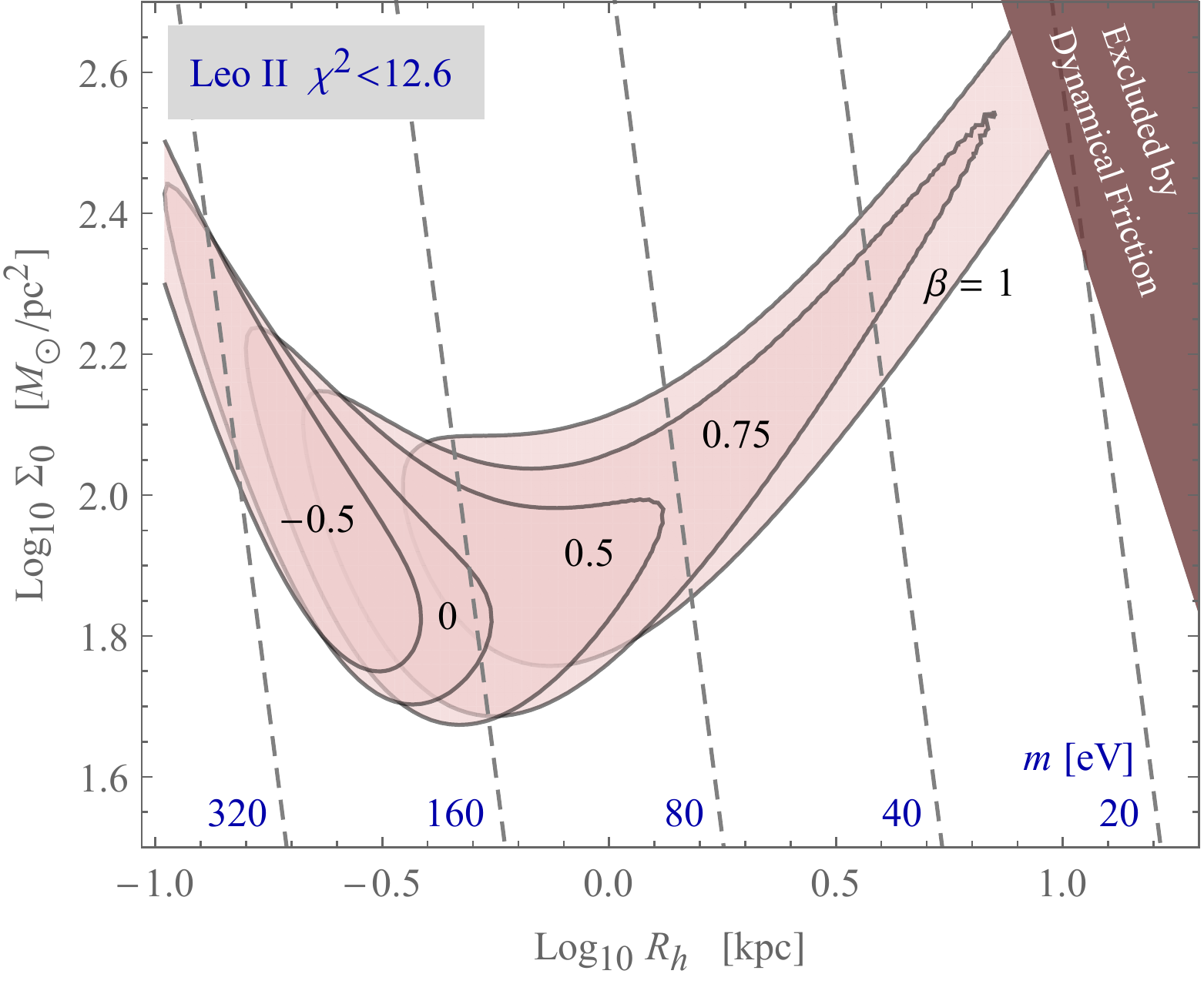}
~~
\includegraphics[width=.95\columnwidth]{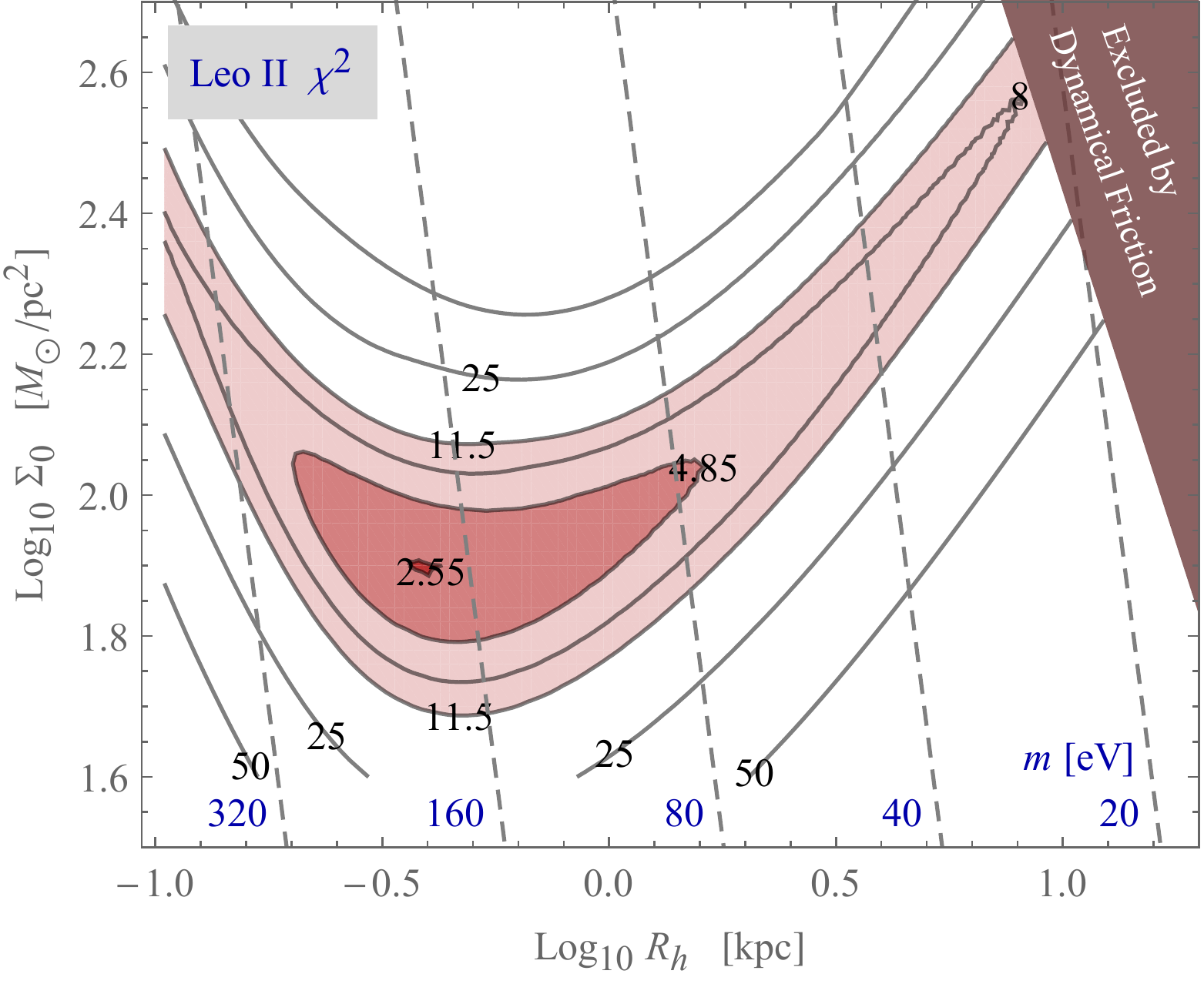}%
\caption{LEO II -- 
  Left: Contours of compatibility with data at 68\% probability
  ($\chi^2\lesssim12.6$, for 11\,dof) as a function of the degenerate DM core parameters, for
  various values of the dispersion anisotropy $\beta$.  
  Right:
  Contours of best $\chi^2$ in the plane $(R_{h},\sigma_0)$ after
  eliminating the anisotropy $\beta$. 
  The light shaded region is compatible with the observational data at
  68\%\,CL ($\chi^2\lesssim11.5$, for 11-1=10\,dof). 
  The darker shading shows the region preferred by data at 68\% CL ($\Delta\chi^2=2.3$), if one {\em
    assumes} the FDM hypothesis as true.}\label{fig_LEOII-results}
\end{figure*}

Our results are shown in Figure~\ref{fig_LEOII-results}.  In the left frame, we plot the $\chi^2$
contours in the plane $(R_{h},\Sigma_0)$, corresponding to 68\% CL exclusion for 11 degrees of
freedom (dof), obtained by assuming fixed values of
$\beta=-0.5,\ldots,1$. 
In the right frame, we eliminate the nuisance parameter $\beta$.
%
As discussed in sec.~\ref{sec:anisotropy}, we use a conservative approach 
that does not require the introduction of a prior on the anisotropy distribution. 
Namely, for fixed model parameters $R_{h}$ and $\Sigma_0$, we minimize the $\chi^2$ over the
admissible range $-\infty \le \beta \le 1$; we then compare the minimal value with a $\chi^2$
probability distribution with $11-1=10\,{\rm dof}$ to possibly reject the assumed parameters.
The light shaded area in the right frame of Figure~\ref{fig_LEOII-results} corresponds to $\chi^2=11.5$ that gives
68\% CL exclusion.\footnote{Because we are interested in a bound on $m$ by
  excluding a region of the $(R_{h},\Sigma_0)$ plane, we use the
  absolute $\chi^2$ to determine the level of compatibility with observational data.  If instead one
  {\it assumes} that the FDM hypothesis is correct, the 68\% CL allowed region in the plane is
  determined by $\Delta \chi^2\equiv \chi^2 -\chi^2_{\rm bf} = 2.3$ (for 2 dof). For completeness,
  we show this region with dark shaded area in
  Figure~\ref{fig_LEOII-results}.}


\begin{figure}
\includegraphics[width=\columnwidth]{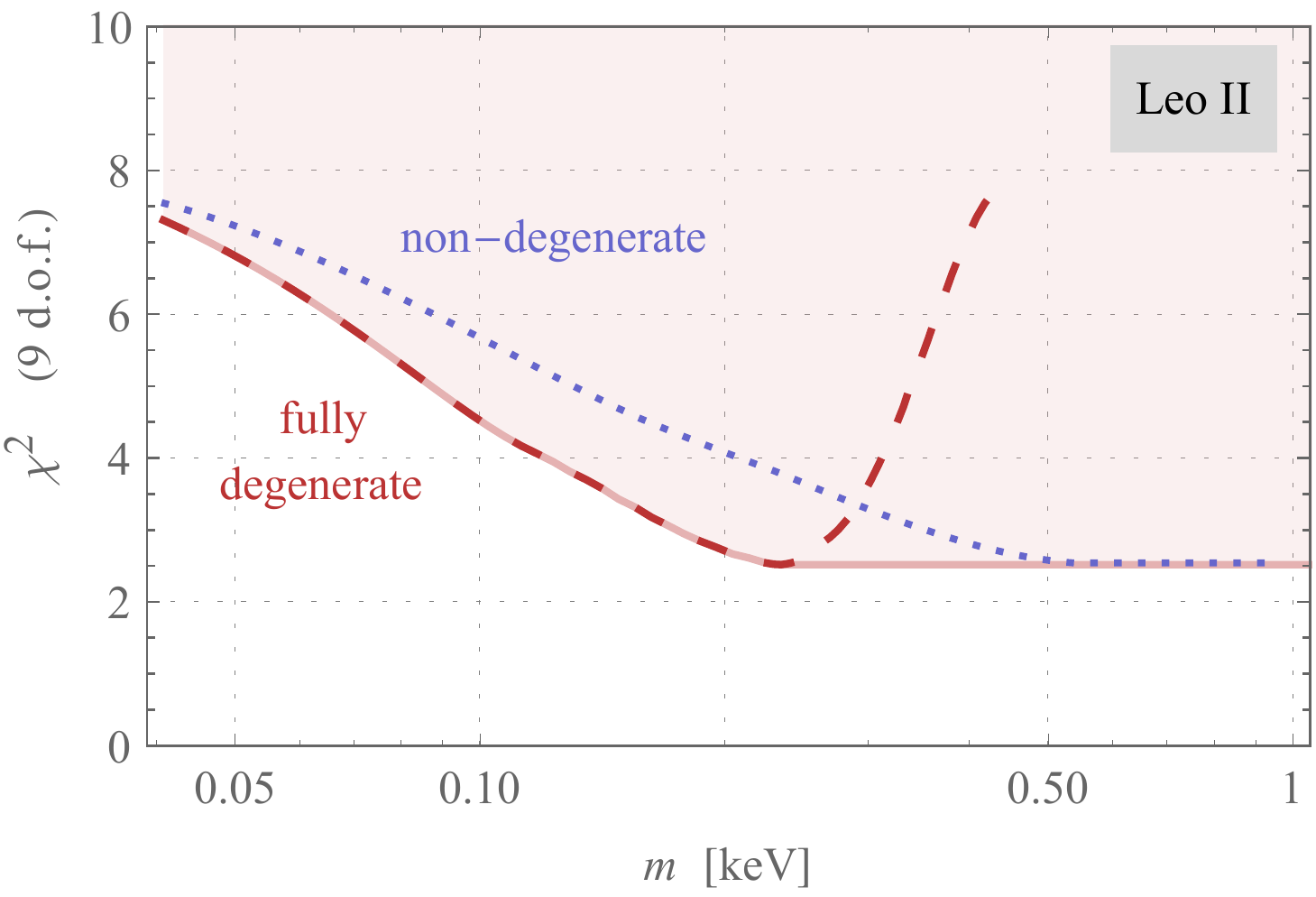}%
\caption{LeoII - The $\chi^2$ as a function of the
particle mass $m$ with different assumptions for the DM distribution.
The red dashed line is obtained, after minimization with respect to $\Sigma_0$ and $\beta$, for
fully degerate cores with radius $R_{h} = R_{\rm h, min}(m,\Sigma_0)$.
The blue dotted line is similarly obtained for a non-degenerate cored (Burkert) profile with radius
$R_{h} = 2\, R_{\rm h, min}(m,\Sigma_0)$.
The light-red solid line and the shaded region represents the overall minimal $\chi^2(m)$, obtained
by considering that DM cores with generic radii $R_{h} \ge R_{\rm h, min}(m,\Sigma_0)$ are
allowed for a fermionic DM candidate.
\label{fig_LEOII-chi2-profile}}
\end{figure}

The best fit model (which has $\chi_{\rm bf}^2\simeq 2.5$) is obtained with $\beta=0.6$ and is
relative to a core size $R_{h}\simeq 400$\,pc and surface density $\Sigma_0\simeq
75\,M_\odot/$pc$^2$.
It provides a very good fit to the observational data, 
as it is shown by the velocity dispersion profile in
figure~\ref{fig_leo_velocities}.  
Being this a fully degenerate halo, the couple of parameters $(R_{h},\Sigma_0)$
corresponds to a well defined mass $m$ of the FDM candidate, which for the best fit corresponds to $m\simeq 0.23$\,keV, in good agreement 
with the value suggested by~\cite{Domcke:2014kla}.

On the other hand, a non degenerate fermionic core can have generic radii $R_{h}$ larger than
the minimal (degenerate) value as found from equation~(\ref{SSigma0})
\be
R_{\rm h, min} =  90\,{\rm
  pc}\,\left(\frac{g}{2}\right)^{-2/5}
       \left(\frac{m}{\rm 1 keV}\right)^{-8/5}
       \left(\frac{\Sigma_0}{\rm M_\odot\, {\rm pc}^{-3}}\right)^{-1/5}
\label{Rhmin2}
\ee
which is shown as gray dashed lines in figure~\ref{fig_LEOII-results} for selected values $m$.
By using the inequality $R_{\rm h, min}\le R_{h}$, one would like to obtain a lower bound on
the DM particle mass directly from the constraints on the halo
parameters $R_{h}$ and $\Sigma_0$, as in equation(\ref{eq:mbound}). 
However, one can see that unless the anisotropy $\beta$ is constrained independently, velocity
dispersion data do not allow to set limits on $m$.
In particular, for maximal values of $\beta\sim1$, i.e.\ radial motion of stars, very small values of $m$ are allowed by the data and are relative to
multi-kpc halos, as it seen in figure~\ref{fig_LEOII-results}.
This situation is most likely not realistic, being unplausible that such a huge DM halo host a
stellar system of just few hundredths pc and small velocity dispersion. It is however not possible to
give at this stage a quantitative support to this commment, using dispersion data alone.  Indeed, in
the upper-right part of the plots, the $\chi^2$ becomes flattish for $R_{h}\gg R_{*}$ because
the limited extent of the stellar component does not permit to constrain a much larger DM halo.

The above conclusion can be made more quantitative by producing a one-dimensional $\chi^2$ profile
as a function of $m$. This is shown in Figure~\ref{fig_LEOII-chi2-profile} which is built as follows.

First, we obtain the red dashed curve, relative to fully degenerate models, by expressing the core
radius ($R_{h}=R_{\rm h, min}$) as a function of the mass $m$ through equation~(\ref{Rhmin2}), and
then minimizing the $\chi^2$ with respect to $\Sigma_0$, in addition to $\beta$.
From this curve one finds directly that no lower limit on the particle mass $m$ can be obtained,
even at 68\% CL that would require $\chi^2>10.4$ (for $11-2=9$ dof). The curve also becomes flattish
for $m\lesssim50\,\eV$, corresponding to the fact that for small masses the DM core becomes much
larger than the stellar component, whose properties can only probe the central density of the DM
distribution, but not its extension.  Incidentally, this also means that the stellar velocity
dispersion can be equally fitted by non-degenerate cored DM profiles with the same central density,
the outer profile being irrelevant.

Then, on the rising part of the red dashed curve, towards large values of $m$, we note that while
the fit worsens progressively as the degenerate core shrinks, one can not set an upper bound on $m$,
because the size of a degenerate core is just a lower limit, and any good fit with a given radius
can also be produced with higher $m$ as a non-degenerate configuration.  Indeed, the mass $m$ has to
play no role in the non-degenerate regime. Thus, the final $\chi^2$ has to be thought as independent
from the DM mass, i.e.\ flat, starting just beyond the minimum of the degenerate case.
This is depicted as a light-red solid line and by the shaded region in
Figure~\ref{fig_LEOII-chi2-profile}.

This argument is confirmed by repeating the whole analisys with a non-degenerate cored density
profile (we adopt the Burkert density profile taking a core radius 2 times larger than the minimal
value, $R_{h} = 2\, R_{\rm h, min}$, see appendix~\ref{app:review}).
This gives the blue dotted curve in Figure~\ref{fig_LEOII-chi2-profile}.  As expected, the
non-degenerate profile leads to better fits for large masses while the $\chi^2$ slowly converges to
the degenerate case at the left end of the plot.  This test could be repeated by considering other
admissible non-degenerate profiles, and/or different $R_{h}>\, R_{\rm h, min}$, and the envelope of
the relative curves will produce the overall minimal $\chi^2$ as depicted as the red solid line.


Even if stellar velocity dispersion data, taken alone, do not allow to obtain a limit on $m$, the
mass of the DM particle can be constrained by considering that the halo decay time due to dynamical
friction~(\ref{tfric}) becomes unacceptably small for very large
objects.
Indeed, in the limit $R_{h}\gg R_{*}$ the quantity directly constrained by stellar velocity
dispersion data is the halo central density $\rho_0 = \Sigma_0 / R_{h}$.
Therefore, by moving along the $\chi^2$ flat direction at increasingly larger radii in
Figure~\ref{fig_LEOII-results}, the halo mass increases as $R_{h}^3$ and eventually reaches values
$M_{h}$ that correspond to unacceptable friction times~(\ref{tfric}).
This constraint is reported on Figs.~\ref{fig_LEOII-results}, by cutting the region where friction
times are unphysical.


In conclusion, the interplay between dynamical friction and velocity dispersion data allows us to
determine an absolute upper bound on the halo size, and thus a lower bound on the DM mass. For
Leo~II, this results in a very weak constraint, $m \gtrsim 25$\,eV.

\subsection{Smallest Dwarf spheroidals: Willman I and Segue I}

\begin{figure}
\centerline{\includegraphics[width=\columnwidth]{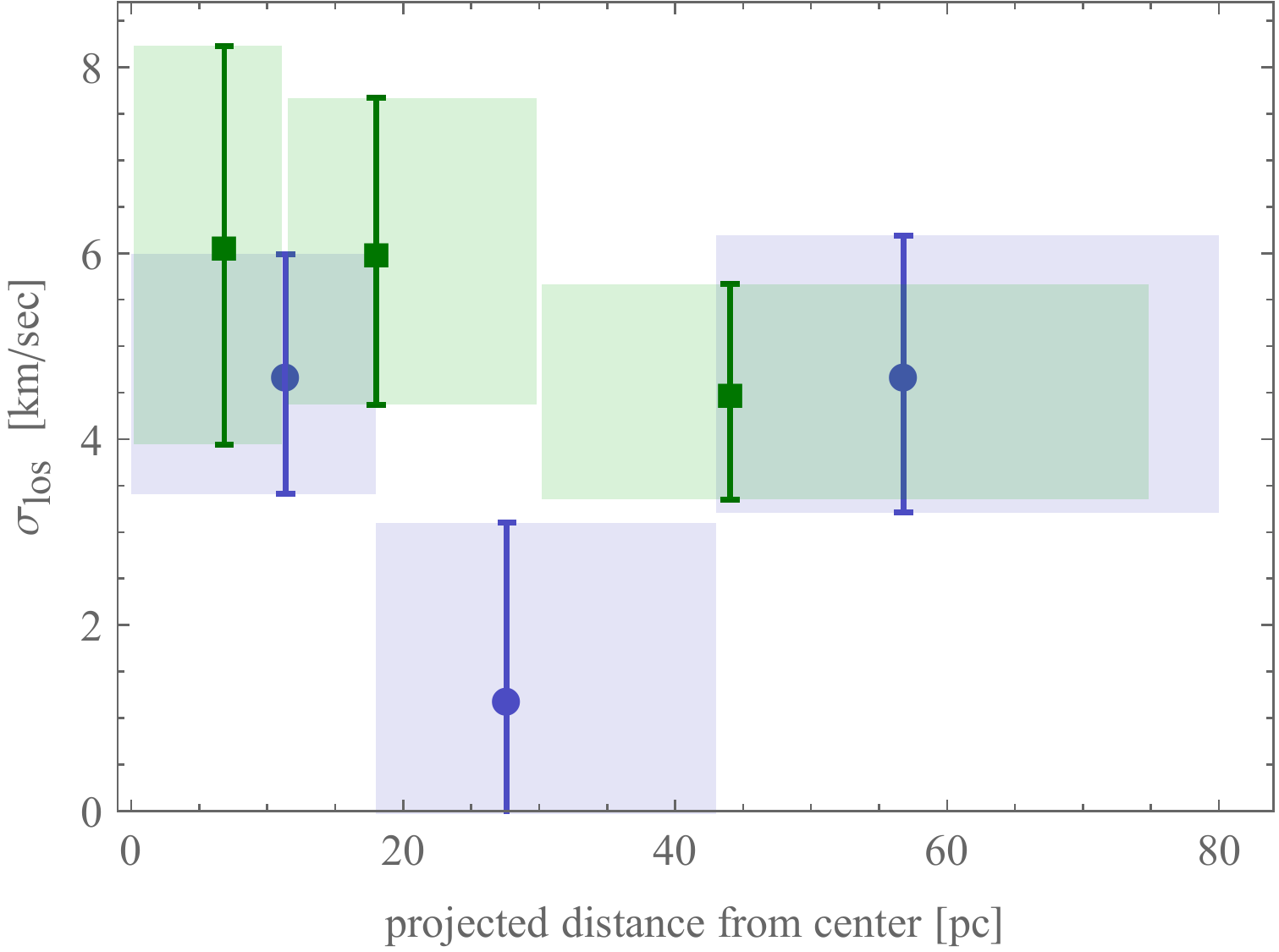}}
  \caption{Line of sight velocity dispersion in Willman I (green squares) and Segue I (blue
    circles).\label{fig_willman_segue_velocities}}
\end{figure}

\noindent
Let us apply the above strategies to the case of the Willman~I and Segue~I dwarf spheroidal
galaxies~\citep{Willman,Simon:2010ek} which are among the smallest galaxies for which line of sight
velocities are available. Their stellar distributions are fitted by Plummer profiles with very small
radii, given by $R_*=25$\,pc and $R_*=29$\,pc, respectively.
We use the stellar velocity data to determine the averaged LOS velocity dispersion,
equation~(\ref{eq:sigmaav1}), in three bins with extension comparable to the Plummer radii.
The results obtained are reported in Figure~\ref{fig_willman_segue_velocities} as a function of the
projected distance from the galactic center.
We compare these observational results with the theoretical predictions by repeating the procedure
adopted for Leo~II, i.e.\ we minimize $\chi^2(R_{h},\Sigma_0,\beta)$ in the full range
$-\infty<\beta\leq1$ of the anisotropy nuisance parameter.

\begin{figure*}
\centerline{\includegraphics[width=0.95\columnwidth]{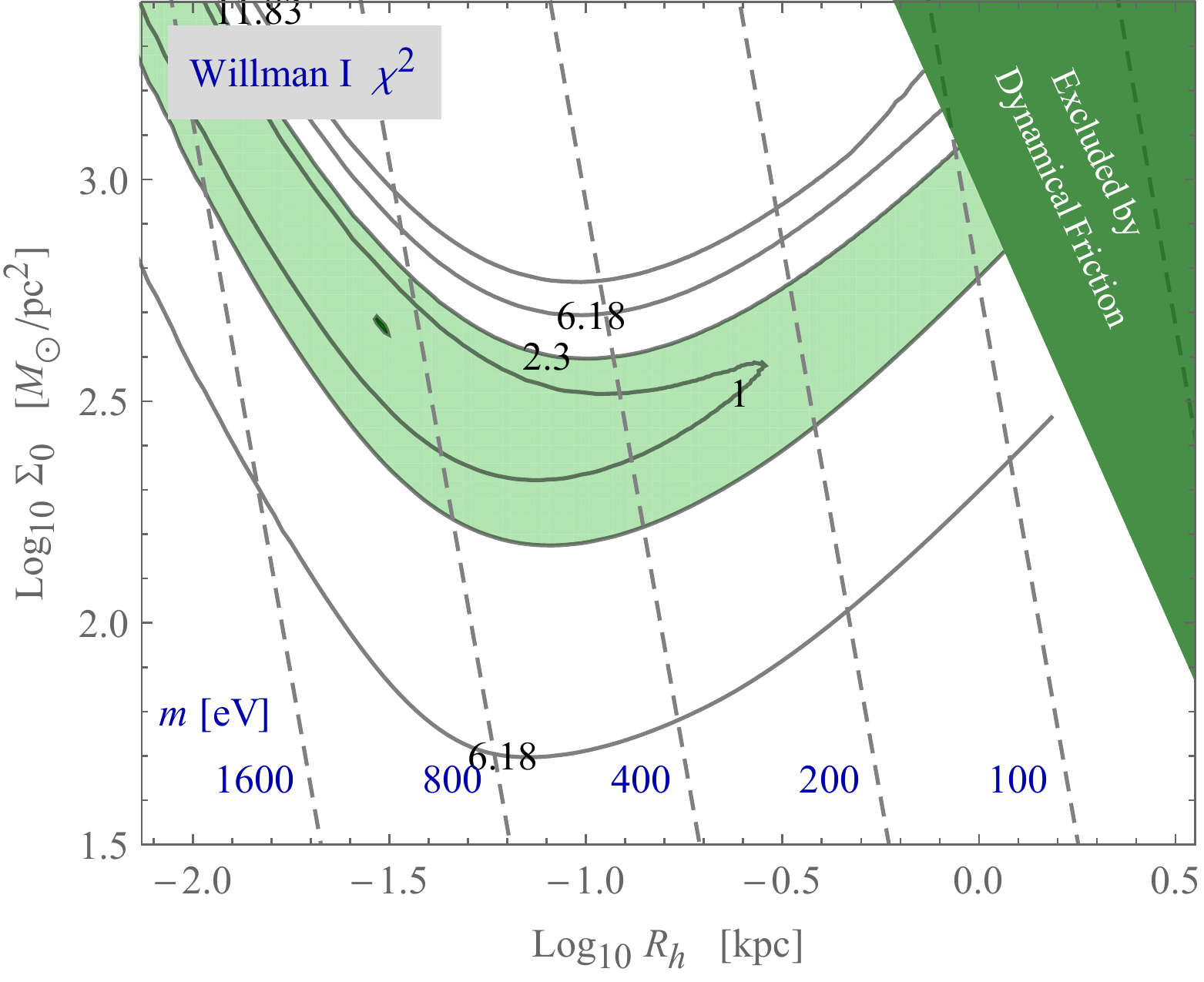}
~~
\includegraphics[width=0.95\columnwidth]{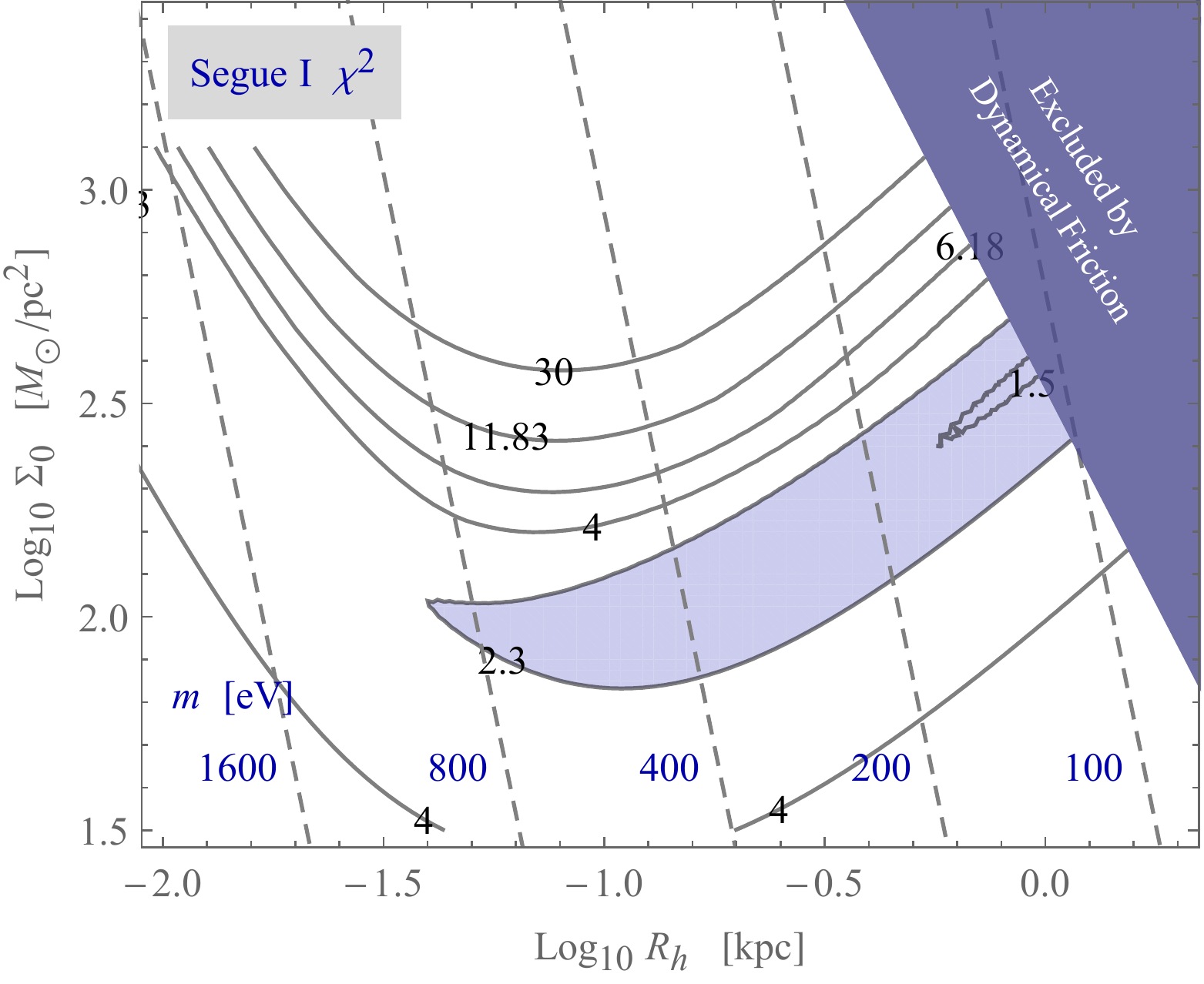}}
\vspace*{-1ex}
\caption{Contours of $\chi^2$, after minimizing on the free range of $\beta$. 
The left panel is for Willmann I, while the right panel refers to Segue
I. The light shaded regions give the region of compatibility with data at 68\% CL  ($\chi^2=2.3$, 2\,dof). 
Notice that no upper limit on the confidence interval on the radius can be placed for either galaxy. 
\label{fig_willman_contours}\vspace*{-1ex}}
\end{figure*}

Our results are reported in figure~\ref{fig_willman_contours}, where we show the contours
corresponding to 68\% CL exclusion for $3-1=2\,{\rm dof}$ for Willman~I in the left frame and for
Segue~I in the right frame ($\chi^2\equiv 2.3$).

We note that no significant constraint on the halo radius is obtained for Willman~I.
Indeed, a good fit is achieved also for $R_{h}\ll R_*$ by allowing the anisotropy
$\beta$ to be increasingly negative.
The fit worsens for large core radii but the $\chi^2$ becomes flattish beyond $R_{h}\sim
300$\,pc, not sufficiently large to give a significative exclusion of larger halo sizes.
This is clearly due to the limited radial extension of the stellar data sample, $\sim 75\,$pc.
The best fit is obtained for $R_{h} \simeq 30\,$pc, $\Sigma_0= 473\,M_\odot/{\rm pc}^2$ and
$\beta = -0.2$ and corresponds to $\chi^2\simeq0$, as it is expected by considering that we have
only three bins and three free parameters.

Again, even if the DM radius is not constrained, the fits provide a useful determination of the
central density of the system, that defines the flat direction in the plane $\left(R_{\rm
    h},\Sigma_0\right)$ for $R_{h}\gg R_{*}$.  Similarly to Leo~II, by requiring that the
dynamical friction decay time for Willman~I is not unphysically small, we can exclude large core
radii beyond $\sim$kpc in the upper-right shaded region in~\ref{fig_willman_contours} and obtain a
robust lower bound on the DM mass, $m\gtrsim 80$\,eV.

A similar analysis holds for Segue I, whose binned velocity dispersions are reported with blue
circles in Figure~\ref{fig_willman_segue_velocities}.
The three bins show that Segue~I presents a different profile with respect to other dwarfs, with the
velocity dispersion rising beyond $\sim40$~pc.  This behavior, that has already been noted in the
literature as a possible evidence of a large DM halo (or an indication against its
regularity/virialization) is consistent with the expectations from a scenario with $R_{h} \gg
R_{*}$.  Consistently, small radii $R_{h}\lesssim 40\,{\rm pc}$ cannot be accommodated by the
fit, even assuming a stellar anisotropy parameter $\beta \le 0$.
The best fit is obtained for $R_{h} \sim {\rm few}\,{\rm kpc}$ and no upper limits can be
derived from Jeans analysis.  Nevertheless, the central halo density is determined and the halo
extension is limited again by the constraint on the dynamical friction time, as shown in
Figure~\ref{fig_willman_contours}.  The resulting lower bound on the DM mass is slightly better to that
derived for Willman~I, $m\gtrsim 100$\,eV, mainly due to the fact that Segue~I is closer to the
galactic center.

\subsection{Other Dwarfs}

\noindent
As a cross check, we note that the results presented above from the Jeans analysis are indeed
compatible with the fits performed in the recent work from~\cite{Hayashi:2016kcy} where the authors
also consider the possibility of triaxial halos and arbitrary density profile slope at the
center. While their analysis does not consider degenerate fermionic halos, their results confirm
that for most dwarf galaxies the DM halo radius is quite poorly constrained, and halos of few kpc
size appear to be allowed by data, even if this is most likely unphysical.  For our purposes, as
discussed above, what drives the bound on the DM mass is the interplay of the dynamical friction
constraint with the central halo density $\rho_0$, which is the quantity constrained by observations
in the limit of large halo size. Because this central density is largely independent on the (outer)
halo shape, we can take advantage of the results of~\cite{Hayashi:2016kcy} to estimate the DM mass
bound for all objects presented in that work.  The bound is in fact driven by the central DM density
and the distance of the dwarf satellite.  As we see, by using the central fitted values of the
densities, our result of Segue I and Leo II are confirmed, so that the results given in this work
represent the most conservative bound around $m\gtrsim100$\,eV, with a possibly slightly stringent
bound from the Triangulum II galaxy.

\begin{table}
\caption{Estimated lower bound on the fermionic DM mass $m$ from  a number of dwarf spheroidal galaxies, adopting the central densities as determined in~\citet{Hayashi:2016kcy}.\vspace*{-1.5ex}}
\label{tab_j}
\renewcommand\arraystretch{0.95}%
\begin{equation*}\begin{array}{|cccc|}
\hline
\text{DSph} &
\begin{array}{c}{\text{Log}}\,\rho_0\\[-.5ex] {[\text{M}_\odot/\text{pc}^3]}\end{array}&
\begin{array}{c} d_0\\[-.5ex]{[\text{kpc}]}\end{array}&
\begin{array}{c}\text{lower bound}\\[-.5ex]\text{on }m\end{array}\\[2.2ex]
\hline
\text{Triangulum II} & 0.3 & 30 & 127 \text{\,eV} \\
 \text{Segue 1} & -0.4 & 32 & 100 \text{\,eV} \\
 \text{Leo T} & -0.6 & 417 & 26 \text{\,eV} \\
 \text{Reticulum II} & -0.8 & 32 & 89 \text{\,eV} \\
 \text{Ursa Major I} & -0.8 & 106 & 49 \text{\,eV} \\
 \text{Coma Berenices} & -0.8 & 44 & 76 \text{\,eV} \\
 \text{Sculptor} & -0.8 & 86 & 54 \text{\,eV} \\
 \text{Fornax} & -1.1 & 147 & 38 \text{\,eV} \\
 \text{Ursa Major II} & -1.2 & 32 & 80 \text{\,eV} \\
 \text{Leo I} & -1.3 & 254 & 27 \text{\,eV} \\
 \text{Canes Venatici II} & -1.4 & 151 & 34 \text{\,eV} \\
 \text{Hercules} & -1.4 & 132 & 37 \text{\,eV} \\
 \text{Pisces II} & -1.5 & 180 & 30 \text{\,eV} \\
 \text{Leo IV} & -1.7 & 158 & 31 \text{\,eV} \\
 \text{Leo II} & -1.7 & 233 & 25 \text{\,eV} \\
 \text{Draco II} & -1.9 & 20 & 82 \text{\,eV} \\
 \text{Sextans} & -2. & 86 & 38 \text{\,eV} \\
 \text{Canes Venatici I} & -2.2 & 224 & 22 \text{\,eV} \\
 \text{Carina} & -2.2 & 106 & 33 \text{\,eV} \\
 \text{Bootes I} & -2.4 & 66 & 39 \text{\,eV} \\
 \text{Leo V} & -2.6 & 178 & 22 \text{\,eV} \\
 \text{Draco} & -2.7 & 76 & 33 \text{\,eV} \\
 \text{Hydra II} & -3.1 & 134 & 22 \text{\,eV} \\
 \text{Segue 2} & -3.2 & 35 & 43 \text{\,eV} \\
\hline
\end{array}\end{equation*}%
\end{table}

\vspace*{-1ex}

\section{Discussion and Conclusions}
\label{sec:conclusions}

\noindent 
In this work we have reassessed the lower bound on the mass of a fermionic dark matter candidate,
independently from particular models of its production or history of its clustering.  The quantum
nature of such light fermionic candidate implies an upper bound on the phase space density in
currently observed objects, and the knowledge of the density can be turned into a lower bound on the
mass, \emph{\`a la} Tremaine-Gunn.  We have reconsidered the smallest Dwarf Spheroidal galaxies,
that according to kinematical data are believed to host the largest densities of dark matter, thus
constituting the ideal candidates to set a lower bound on the DM mass $m$.

Such a bound must be set in the hypothesis that the DM halo of some of these objects is composed by
a completely degenerate gas of fermions, whose density profile is defined by the Lane-Emden
equation. We have performed a fit of the stellar velocity dispersion predicted by the gravitational
potential generated by such DM halo, versus the observed stellar dispersion velocity and density
profile of the Willman I, Segue I and Leo II galaxies.
In our analysis, differently from recent works on the subject, we have not assumed that luminous
matter traces the DM distribution, thus we have considered the DM core radius and surface density as
free parameters.
Moreover, we have taken into account the effect of the unknown anisotropy of the stellar velocity
dispersion and marginalized over it.

As we have shown, the nuisance due to the stellar velocity anisotropy $\beta$ seriously hampers the
possibility to efficiently constrain the DM halo parameters.  In practice, one finds equally
acceptable halos of very small sizes and negative $\beta$ (where the total DM halo mass is
determined) or very large sizes \emph{$\sim$ few kpc} and anisotropy near 1 (in which case the inner
DM spatial density is determined).  This latter scenario effectively corresponds to low phase-space
densities, and thus no sensible lower bound on $m$ can be given from stellar kinematical data
alone. This situation is likely to persist even in the future, until a way to measure the velocity
anisotropy in dwarfs spheroidals will be available (see e.g.~\cite{2017arXiv170104833R}) although
this appears currently quite unconceivable.

New approaches have been proposed to circunvent the $\beta$-degeneracy in dwarf spheroidals in which
sub-populations can be separated~\citep{Battaglia:2008jz,Walker:2011zu,Agnello:2012uc}.
These methods were applied to the Fornax galaxy in~\cite{Amorisco:2012rd} to exclude the NFW profile
and constrain the DM distribution.
The upper bound on Fornax core radius was used by~\cite{Randall:2016bqw} to infer a lower limit $m
> 70$\,eV for the mass of a fermionic dark matter particle.
Unfortunately, the likelihood distribution used to limit the core size in \cite{Amorisco:2012rd}
does not converge to zero for large core radii (see their Figure~4), as it is expected due to the
fact that the stellar populations have limited extent.
Therefore, while providing a robust lower bound for the core radius, even this approach cannot
exclude few-kpc core radii at high confidence level.

Such multi-kpc halos are in any case unrealistic and a rationale to rule them out is
provided~\cite{1992ApJ...389L...9G} by the fact that very large halos of known density correspond to
large total halo mass, which makes their time of orbital decay due to dynamical friction in the
Galactic DM halo, formula~(\ref{tfric}), unphysically small.  Therefore, dynamical friction can be
used to effectively limit the halo size and the interplay with the quantum bound on phase-space
density leads finally to a lower bound on the fermionic DM mass~$m$.

As it turns out from the analysis that we described, at present the most restrictive bound stems
from the study of the Willman I and Segue I galaxies.  Our results are put together in
Figure~\ref{fig_bananas}, where only the interplay between the fit to stellar data and the
constraint from dynamical friction leads to a robust lower bound of $m\gtrsim 100$\,eV.  Thus, one
is led to reopen the case for sub-keV fermionic DM, like sterile neutrinos of mass down to 100\,eV.

For these two small dwarf galaxies driving the bound, the resulting DM halo can reach sizes of
$\sim1$\,kpc, much larger than their stellar components.  This does not mean that all the dwarf
spheroidal galaxies shall have such enhanced halos; this could likely hold only for these smallest
objects that approach the fermionic degenerate regime.

As far as DM indirect detection is concerned, we note that the expected flux from DM annihilation
(so called $J$-factor) is enhanced in the limiting case of the extended halo sizes considered here,
compensating the naturally low flux characteristic of cored halos. At the same time, the dynamical
friction upper bound on the halo sizes will slightly reduce the maximal expected $J$-factor in cored
halos, with respect to the analysis of e.g.~\cite{Hayashi:2016kcy}.

Clearly, DM masses $m=100\,$eV are at odds with bounds derived from the effect of warm dark matter
on structure formation (e.g.\ Lyman-$\alpha$) that typically forbid masses below few\,keV, see
e.g.~\cite{Irsic:2017ixq}, by limiting their free streaming length.  Therefore, for this scenario to
be realistic, the spectrum of such DM candidates should be much colder than usual (see
e.g.~\cite{Adhikari:2016bei}).  This can be realized in models with production via decay as e.g.\
in~\cite{Petraki:2007gq,Domcke:2014kla}, or for instance in models in which DM at decoupling is
overabundant and then subject to dilution by decays of other species, along the lines of
e.g.~\cite{Bezrukov:2009th,Nemevsek:2012cd}.

More theoretically, in order to attain the fermionic degeneracy that we have tested, it is also
necessary that either the maximum of the primordial phase-space density saturates the occupation
limit (\ref{WDM_flimit})\ as it happens e.g.\ in relativistic decoupling, or alternatively that DM is
subject to some form of dissipation or interaction, so that the phase-space density might grow
during collapse.  Indeed, a very interesting (and outstanding) issue is that of which collapse
mechanism and time scales could lead to degenerate fermionic halos.  While the free energy and
entropy budget have been shown to be favorable~\citep{Hertel:1972fp,Bilic:1996mv, Chavanis:2002rj},
an assessment of the dynamics and relaxation times is still beyond reach, see~\cite{Chavanis:2014woa,
  Campa:2009jxa}.

On the observational side, it is worth commenting that while the DSph galaxies are the smallest and
most DM dominated astrophysical objects, with a number of new dwarfs being currently discovered by
present surveys, the possibility of using other types of galaxies for setting a bound on $m$ from
degeneracy is also of interest.  Recently, cored halo mass modelings of Disk Dwarf galaxies from the
Little Things survey have been performed, see~\cite{Karukes:2016eiz}. Although the rotation curve
decomposition is affected by uncertainty in the asymmetric drift gas contribution, due to their disk
structure they are not subject to the dramatic anisotropy nuisance parameter of dwarf spheroidals
and could potentially lead to a better bound on the DM mass $m$.

\vspace*{-2em}

\section*{acknowledgments}

\noindent 
We thank Neven Bili\'c, Kathy Karukes, Paolo Salucci and Piero Ullio for useful discussions.
F.N. was partially supported by the H2020 CSA Twinning project No.~692194 ``RBI-T-WINNING''.
Results of this work were also presented in
\href{https://indico.cern.ch/event/505065/contributions/2166456/}{https://indico.cern.ch/event/505065/contributions/2166456/}.


\begin{appendix}
\section{Self-gravitating fermionic gas}

\label{app:review}
\noindent 
We briefly review in this appendix the analysis of the equilibrium distribution of a
self-gravitating gas of neutral fermions. We describe first the limiting case of complete
degeneration and then we recall the Thomas-Fermi treatment which allows to describe the transition
to partially or non-degenerate case.

\paragraph{Stability conditions for the DM halo.} 
If we have large number of DM particles, we can assume they move in a spherically symmetric
mean-field gravitational potential $\phi (r)$ which satisfies the Poisson's equation:
\begin{eqnarray}
\label{WDM_form5}
\frac{d\phi}{dr}&=& \frac{G M}{r^2} \nonumber \\
\frac{dM}{dr}&=& 4\pi r^2 \rho \,,
\end{eqnarray} 
where $M(r)$ is the mass enclosed within the radius $r$, $G$ is the Newton's constant and $\rho(r)$
is the matter density.
For non relativistic particles, the density can be expressed as:
\begin{equation}
\label{WDM_form7}
\rho = m \int dp\;  4\pi p^2 \, f(p)\,,
\end{equation} 
where $m$ is the particle mass and we assumed that DM distribution function $f(p)$ is isotropic.
The dynamical stability of the system is expressed by the Jeans equation:
\begin{equation}
\frac{d}{dr} (\rho \sigma_{\rm DM}^2) =  - \rho \frac{d\phi}{dr},
\label{WDM-Jeans}
\end{equation}
where the DM velocity dispersion $\sigma^2_{\rm DM}$ is given by:
\begin{equation}
\sigma_{\rm DM}^2 = \frac{1}{3}\frac{\int  dp\; (p^4/m^2) \, f(p)}{\int  dp\;  p^2 \, f(p)}
\end{equation}
If DM is composed by fermions, the distribution function $f(p)$ has an upper limit:
\begin{equation}
f(p) \le \frac{g}{(2\pi  \hbar)^3},
\label{WDM_flimit}
\end{equation} 
where $g$ represents the number of internal (spin) degrees of freedom. This automatically implies
that a lower limit exists for the velocity dispersion
\begin{equation}
\sigma_{\rm DM}^2\ge \sigma^2_{\rm DM, min} = \frac{1}{5} 
\left(\frac{ 6\, \pi^2 \hbar^3 \rho}{g\, m^4}\right)^{2/3}
\end{equation}
of a fermionic system of fixed density.

\medskip

\paragraph{The strong degeneracy limit.}
In the strong degeneracy regime, the states with energy below the Fermi energy $\varepsilon$ are
fully occupied, i.e. the distribution function $f(p)$ has the form:
\begin{equation}
f(p)=\left\{\begin{array}{ll}  \frac{g}{(2\pi \hbar)^3} &\qquad p < p_{\rm
                                                          F}
                    \\0&\qquad p > p_{\rm F}\end{array}\right.
\label{eq:fDM}
\end{equation}
where $p_{\rm F} = \sqrt{2 m \varepsilon}$ is the Fermi momentum.
In this assumption, one obtains the expressions:
\begin{eqnarray}
\nonumber
\rho &=&  K \, \varepsilon^{3/2}\\
\label{rhosigmadeg}
\sigma_{\rm DM}^2 &=& K'  \,\varepsilon\,,
\end{eqnarray}
where $K= \sqrt{2} \, g\, m^{5/2} / (3 \pi^2 \hbar^3)$ and $K' = 2/(5m)$, so that 
Eqs.~(\ref{WDM_form5},\ref{WDM-Jeans}) can be recasted in the form
\begin{equation}
\frac{1}{r^2}\frac{d}{dr}\left[r^2\frac{d\varepsilon(r)}{dr}\right] =
- 4\pi G m \, K \varepsilon(r)^{3/2}\,.
\label{SDeps}
\end{equation}
This equation has to be integrated with the condition $d\varepsilon(0)/dr = 0$ ensuring that
the gravitational acceleration is zero at the center. By defining
\begin{equation}
\xi \equiv r/\tilde{r} \;, 
\end{equation}
where the scale radius $\tilde{r}$ is given by
\begin{equation}
\tilde{r}\equiv \frac{1}{\sqrt{4 \pi G
m \, K\varepsilon_0^{1/2}}} = \frac{1}{\sqrt{4 \pi G
m \, K^{2/3} \rho_0^{1/3}}} ,
\end{equation}
and by using the function $\theta(\xi)$ defined as
\begin{equation}
\theta(\xi) \equiv \frac{\varepsilon(\xi)}{\varepsilon_0} =
\left[\frac{\rho(\xi)}{\rho_0}\right]^{2/3} \, ,
\label{FDM-ScaleRadius}
\end{equation}
where $\varepsilon_0$ ($\rho_0$) is the central value of the Fermi
energy (density), equation~(\ref{SDeps}) can be rewritten in the form
\begin{equation}
\frac{1}{\xi^2}\frac{d}{d\xi}\left[\xi^2\frac{d\theta(\xi)}{d\xi}\right]
= -\theta(\xi)^{3/2} \,,
\label{FDM-LaneEmden}
\end{equation}
which is the well-know Lane-Emden equation.

\begin{figure}
\vspace*{-1ex}
\centerline{\includegraphics[width=\columnwidth]{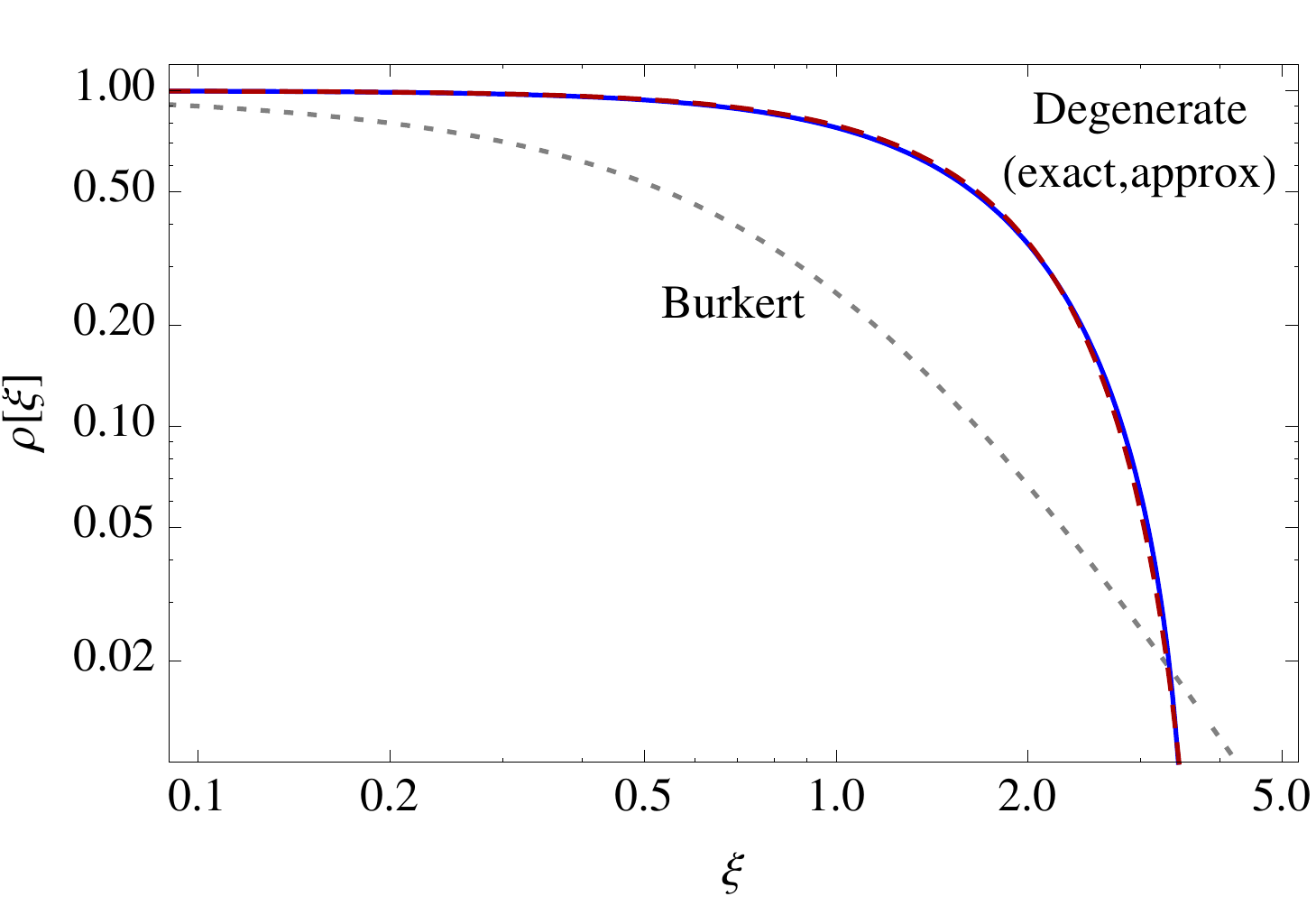}}
\vspace*{0ex}
\caption{Density profiles for the DM halos. Solid: the solution of the Lane-Emden equation for
  degenerate fermions. Dashed: its approximation adopted in the text~(\ref{eq:rhoDM}). Dotted: the
  Burkert density profile for comparison.\label{fig:profiles}}
\end{figure}

Eqs.~(\ref{FDM-ScaleRadius}) and (\ref{FDM-LaneEmden}) show that the profiles of degenerate
Fermionic DM halos are universal and depend only on the assumed central density $\rho_0$ and DM
particle mass $m$. The mass distribution crucially differs from the usually adopted cusped or cored
profiles.  A sharp transition exists from an internal core with quite uniform density to an external
region devoid of DM.  For our purposes, the density profile of degenerate fermionic DM halos can be
well approximated by
\begin{equation}
\rho(\xi)=\rho_0 \cos^3 \left[\frac{\pi}{8} \,\xi\right]
\label{eq:rhoDM}
\end{equation}
for $0\le \xi\le 3.65$, and $\rho(\xi)=0$ elsewhere, see Figure~\ref{fig:profiles}.


We define the halo radius by the condition $\rho(\xi_{h}) = \rho_0/4$ 
that gives $\xi_h = 2.26$ corresponding to:
\begin{eqnarray}
\label{RhAppend}
R_{h} &\equiv& \xi_{h} \tilde{r} = \\
\nonumber
&  = & 2.26  \left(\frac{9 \pi}{2^7} \right) ^{\frac{1}{6}}\frac{\hbar}{G^{1/2}} \;\;\; g^{-\frac{1}{3}}m^{-\frac{4}{3}}\rho_0^{-\frac{1}{6}}= \\
\nonumber
& = & 42.4 \, {\rm pc}\, \left(\frac{g}{2}\right)^{-1/3}
\left(\frac{m}{\rm 1 keV}\right)^{-4/3}
\left(\frac{\rho_0}{\rm M_\odot/pc^{3}}\right)^{-1/6}\,.
\end{eqnarray}
By using the above expression, we rewrite equation~(\ref{eq:rhoDM}) in the form:
\begin{equation}
\rho(r)=\rho_0 \, \cos^3 \left[\frac{25}{88}\, \pi\, 
\frac{r}{R_{h}}\right],
\end{equation}
where we used the approximate equality $\xi_{h} / 8 \simeq 25/88$.

\paragraph{The Thomas-Fermi model for fermionic DM.}

A self consistent description of isothermal fermionic DM halos with an arbitrary level of
degeneration can be obtained by using a Thomas-Fermi approach,
see~\cite{Bilic:2001es,deVega:2013jfy}.  One assumes that DM particles follow a Fermi-Dirac
distribution
\begin{equation}
\label{WDM_form14}
f_{\rm FD}(p;T,\mu)= \frac{g}{(2\pi \hbar)^3}\frac{1}{\exp[(E-\mu)/T]+1} \,,
\end{equation} 
where $E=p^2/(2m)$ is the single-particle kinetic energy, $T$ is the temperature expressed in terms
of energy and $\mu$ is the chemical potential.  In the above, the density can be expressed as:
\begin{equation}
\rho = \frac{g \,(2T)^{3/2}m^{5/2}}{6\pi^2 \hbar ^3}\, I_{2}(\nu)\,,
\end{equation}
where
\begin{equation}
I_{2}(\nu)\equiv 3 \int_{0}^{\infty} dy \, \frac{y^2}{\exp{(y^2-\nu)}+1}
\end{equation}
and $\nu\equiv \mu/T$ is a degeneracy parameter.

If one assumes a constant temperature, $T(r)\equiv\tilde{T}$, and that the chemical
potential at each given radius includes the gravitational potential $\phi (r)$ as
\begin{equation}
\label{WDM_form6}
\mu(r) = \tilde{\mu} -m\phi(r) 
\end{equation} 
where $\tilde{\mu}$ is a constant, then the Jeans equation (\ref{WDM-Jeans}) is
automatically fulfilled and Eqs.~(\ref{WDM_form5}), (\ref{WDM_form7}) can be recast in the form
\begin{equation}
\frac{1}{r^2}\frac{d}{dr}\left[r^2\frac{d\mu(r)}{dr}\right] = - 4\pi G m
\rho(r) \,,
\label{Eq:ThomasFermi}
\end{equation}
again to be integrated with the condition $d\mu(0)/dr = 0$
for zero gravitational acceleration at the galaxy center.

\paragraph{The non degenerate case.} The Thomas-Fermi approach just described has the advantage of
automatically implementing the upper limit (\ref{WDM_flimit})\ imposed by the Pauli exclusion
principle; it can thus describe the transition between classical and degenerate structures, in a
continuous way. By using this approach, one is able to see that when $R_{h}$ is 2-3 times larger
than the minimal value in equation~(\ref{RhAppend}) the fermionic nature of DM particles can be
neglected, i.e.\ the resulting structures are essentially indistinguishable from cored isothermal
halos obtained by assuming Maxwell-Boltzmann statics and arbitrary values of the particle mass $m$.

This approach does not allow, however, to unambiguously predict the halo properties in the non
degenerate case.
Indeed, in the classical regime (i.e.\ for $\nu\ll 1$) one has $I_2(\nu)\simeq\exp(\nu)$, differently
from the strongly degenerate case in which $I_2(\nu)\simeq \nu^{3/2}$.
Thus, the r.h.s.\ of equation~(\ref{Eq:ThomasFermi})\ depends both on the temperature and the chemical
potential. One obtains then a family of solutions depending on these two free parameters: the
temperature $\tilde{T}$ and the assumed chemical potential $\mu_0$ (or, equivalently, the assumed
density $\rho_0$) at the center of the system.
Morevoer, a temperature profile, here constant, had to be assumed in the Thomas-Fermi approach in
order to solve equation~(\ref{WDM-Jeans}).
In a more realistic scenario, in which the temperature may vary along the galactic structure,
$\tilde{T}$ could be regarded as the central temperature; the predictions obtained in the degenerate
or semi-degenerate regimes are thus valid in the central core where temperature variations can be
neglected, while the properties of the external region depend on the radial temperature profile. As
it is natural to expect, basing on sole theoretical grounds it is thus impossible to predict the
mass distribution in regions of non degeneration.

For this reason in the text, where we refer to non-degenerate halos, we model them by using the
observationally supported Burkert profile:
\begin{equation}
\rho_{\rm Bur}(r) = \frac{\rho_0}{(1+x)(1 + x ^2)} \,,\qquad x= r/R_{h}\,.
\label{eq:Burkert}
\end{equation}
In using this profile, we require that the central density $\rho_0$ and core radius $R_{h}$ are
consistent with the assumption of non degenerate structure composed by fermions with mass $m$ and
$g$ spin degrees of freedom, i.e.\ for each assumed value of $\rho_0$, the halo radius $R_{h}$
is required to be a factor $\sim 2$ larger than the degenerate limit expressed by
equation~(\ref{RhAppend}).

\end{appendix}

\bibliography{Bibliography1}

\bsp	
\label{lastpage}
\end{document}